\documentclass{jfm}
\usepackage{amsmath,cleveref,bm}
\usepackage{mathtools}
\usepackage{graphicx}
\graphicspath{{Figures/},{Figures_v2/},{Figures_SS_v3/}}
\usepackage{afterpage}
\usepackage{enumitem}
\usepackage{xcolor}
 \usepackage{natbib}
 \usepackage{appendix}
 \usepackage{makecell}
\usepackage{subfig}
 \usepackage[utf8x]{inputenc}
\usepackage[T1,T2A]{fontenc} 
 \usepackage[main=english,russian]{babel}
 \usepackage{breqn}
 \usepackage{placeins}
\usepackage{lineno}
\usepackage{upgreek}
\usepackage{epstopdf}
\usepackage{pdfpages}

\usepackage{verbatim}

\makeatletter
\newcommand{\specialnumber}[1]{%
	\def\tagform@##1{\maketag@@@{(\ignorespaces##1\unskip\@@italiccorr#1)}}}
\newcommand{\specialeqref}[2]{\begingroup
	\def\tagform@##1{\maketag@@@{(\ignorespaces##1\unskip\@@italiccorr#2)}}%
	\eqref{#1}\endgroup}

\title[The increased wave-induced drift of floating marine litter]{A mechanism for the increased wave-induced drift of floating marine litter}
\author[Calvert et al.]{R. Calvert$^{a,b}$, M.L. McAllister$^{a}$, C. Whittaker$^{c}$, A. Raby$^{d}$, A.G.L. Borthwick$^b$ and T.S. van den Bremer$^{a,e}$}
\affiliation{
$^a$Department of Engineering Science, University of Oxford, Oxford OX1 3PJ, UK\\[\affilskip]
$^b$School of Engineering, The University of Edinburgh, Edinburgh, EH9 3FB,  UK\\[\affilskip]

$^{c}$Department of Civil and Environmental Engineering, University of Auckland, Auckland 1010, New Zealand\\[\affilskip]
$^{d}$School of Engineering, University of Plymouth, Plymouth PL4 8AA, UK\\[\affilskip]
$^{e}$Faculty of Civil Engineering and Geosciences, Delft University of Technology, 2628 CD, Delft, The Netherlands\\[\affilskip]
}

\begin{document}

\maketitle
\begin{abstract}
Periodic water waves generate Stokes drift as manifest from the orbits of Lagrangian particles not fully closing. Stokes drift can contribute to the transport of floating marine litter, including plastic. Previously, marine litter objects have been considered to be perfect Lagrangian tracers, travelling with the Stokes drift of the waves. However, floating marine litter objects have large ranges of sizes and densities, which potentially result in different rates of transport by waves due to the non-Lagrangian behaviour of the objects. Through a combination of theory and experiments for idealised spherical objects in deep-water waves, we show that different objects are transported at different rates depending on their size and density, and that larger buoyant objects can have increased drift compared with Lagrangian tracers. We show that the mechanism for the increased drift observed in our experiments comprises the variable submergence and the corresponding dynamic buoyancy force components in a direction perpendicular to the local water surface. This leads to an amplification of the drift of these objects compared to the Stokes drift when averaged over the wave cycle. Using an expansion in wave steepness, we derive a closed-form approximation for this increased drift, which can be included in ocean-scale models of marine litter transport.
\end{abstract}

\section{Introduction}
In the last half century large concentrations of plastic have polluted the oceans, with harmful effects on marine wildlife and potentially on human health \citep{ostle2019,cozar2014,cole2011}. Plastic pollution may have lasting impact, noting that it has been estimated that plastic may take hundreds or thousands of years for plastic to decay in the ocean \citep{cole2011}, although such estimates are subject to considerable uncertainty \citep{ward2020opinion}. Floating plastic debris is transported and dispersed by three key mechanisms: currents, wind, and waves \citep{vansebille2020}. This paper will investigate wave-induced transport.

To leading order and in deep water, the Lagrangian motion induced by waves takes the form of circular orbits with Lagrangian particles following these orbits in a periodic fashion. The imbalance between the forward orbital velocity when under the crest and backward orbital velocity when under the trough, caused by the decay in velocity with depth, and the fact that particles spend more time under the forward-moving crest than under the backward-moving trough results in orbits that do not close, i.e. a Lagrangian-mean drift, known as Stokes drift \citep{stokes1847}. Stokes drift in deep water is proportional to the square of wave steepness and decays with depth at twice the rate of the oscillatory water particle velocity (see e.g. the review by \citet{vandenbremer_breivik2017}). Ocean surface gravity waves are driven by wind, and thus Stokes drift has often been assumed to be locally proportional to the wind forcing \citep{weber1983}. However, waves are slow to build and, once established as swell, waves can travel long distances with little dispersion \citep{ardhuin2019,hanley2010}, and so their magnitude is not always proportional to the local wind forcing. Wave models, such as WaveWatch III \citep{tol:MMABmanual}, can be used to predict Stokes drift \citep{webb_etal2011,webb_foxkemper2015}.

Several authors have considered the effect of Stokes drift on the transport of floating marine litter. In an early study, \citet{kubota1994} found that Stokes drift derived from local wind fields did not make a significant contribution towards debris transport. However, more recent studies that included the entire wave field showed that Stokes drift could play an important role. For example, \Citet{iwasaki2017} found that Stokes drift transported plastic towards the coast in the Sea of Japan during winter, and  \citet{delandmeter2019} reported similar behaviour in the Norwegian Sea. Stokes drift could enable debris to leak out of the Indian Ocean \citep{dobler2019}, cause drifting debris to cross the strong circumpolar winds and currents to reach the Antarctic coast \citep{fraser2018}, and thus promote increased transport to polar regions \citep{onink2019}. \citet{isobe2014} modelled the plastic beaching process by including Stokes drift and sinking velocity and observed that larger plastic debris was selectively moved onshore. All the foregoing studies have simply assumed that floating marine litter objects are transported with the Stokes drift; in other words, that they are perfect Lagrangian tracers.

If a particle is infinitesimally small and has the same density as water, it will behave purely as a Lagrangian tracer and will be transported with the Stokes drift. This is not necessarily true for an object of finite size or of a density different to that of water. As the inertia of such an object becomes important, the fluid will exert a drag on the object owing to the relative velocity between the object and fluid. Furthermore, the object may rise, sink, or float depending on the density difference. The literature distinguishes between fully submerged and floating objects, discussed separately below. 

The motion of a fully submerged sphere in unsteady flow with viscous drag can be described by the Maxey--Riley equations \citep{maxey1983}. Based on this pioneering work, \citet{eames2008} and \cite{santamaria_etal2013} examined how far slightly positively or negatively buoyant objects would be transported by regular waves. They defined the distance transported as either the horizontal distance transported whilst a negatively buoyant object sinks from the free surface to the sea floor or the horizontal distance transported whilst a positively buoyant object rises from the sea floor to the free surface. \cite{eames2008} and \cite{santamaria_etal2013} used an expansion in wave steepness and Stokes number to arrive at analytical solutions for small objects. To leading order and for negatively buoyant objects, \cite{eames2008} showed such small objects are transported with a mean horizontal Stokes drift velocity and sediment with their terminal fall velocity. \cite{santamaria_etal2013} predicted that positively buoyant objects would experience an increase in drift owing to their inertia. Although \citet{eames2008} and \citet{santamaria_etal2013} considered the object's inertia when examining transport by waves, both considered completely submerged objects. 

Also considering fully submerged objects, \citet{dibenedetto_ouellette2018} first showed non-spherical objects have a preferential orientation under waves, confirming this result numerically \citep{dibenedetto_ouellette2018} and experimentally \citep{Dibenedeto2019Oreintation} but not examining the effect of the object's inertia. The orientation changes the drag on slightly negatively buoyant objects, which results in objects of different shapes being transported different distances before `raining out'  \citep{dibenedetto_etal2018}. 

Analysis of the motion of floating objects commences with the extension of Maxey--Riley equation \citep{maxey1983} to include a free surface as undertaken by \citet{rumer1979}. These authors considered the free surface to be an oscillating slope with a vertical force balance between gravity and buoyancy, whilst the horizontal part of the buoyancy force induces object motion in what \citet{rumer1979} termed the slope-sliding effect. \Citet{shen2001} further extended the slope-sliding model, proceeding to find analytical solutions of the object motion in limit of no added mass or no resistance. \Citet{huang2016} found that the drift of relatively large floating discs, used to model floating ice sheets, increased beyond the Stokes drift in physical experiments. This could be explained by numerical solutions to an equation of motion based on a rotating coordinate system which aligned with the free surface, leaving the physical mechanism at work unclear. 

Although not focusing on waves, \citet{beron2016inertia} showed that the inertia of an undrogued drifter is important for their accumulation in subtropic gyres. The study integrated a Maxey--Riley equation that modelled the variable submergence of surface drifters and included forcing from current and wind velocities, by varying the relative effect of each with the submerged volume of the drifter. The drag formulation assumed linear dependence of force on the density ratio between the object and water, as has been experimentally validated by \citet{miron2020laboratory}. The Maxey--Riley equation has been extended to model floating Sargassum rafts \citep{beron-vera_miron_2020}.

Surface tension can be important in the response of small inertial particles under wave action, as shown by \citeauthor{falkovich2005floater} (\citeyear{falkovich2005floater}), who found that hydrophobic and hydrophilic particles concentrate in antinodes and nodes of a standing wave, respectively. \citet{Denissenko2006How} demonstrated the importance of surface tension when predicting time scales of small particle clusters in standing waves. In this paper, we do not examine the effect of surface tension, which places a lower limit on the size of particles for which our model is valid.

This paper examines the transport of inertial, finite-size floating marine litter under the influence of non-breaking waves. Our derivation starts from Newton's second law, with buoyancy, gravity and drag force components. Using a transformed coordinate system, similar but not equivalent to \citet{huang2016}, that vertically translates and is oriented orthogonally to the time-varying free surface, we ensure that the dynamic buoyancy term is directed normal to the free surface. In this model, the drag force changes with submergence of the object, and we formulate a drag coefficient that is valid across a range of Reynolds numbers. We use perturbation methods to derive a closed-form solution for the transport of inertial, finite-size floating spherical objects, which is then used to interpret the physical mechanism for their enhanced transport compared to the Stokes drift. Numerical and analytical solutions are compared for viscous drag. In order to observe the predicted response, we perform experiments in a laboratory wave flume. 

This paper is laid out as follows. \S \ref{s:maths} presents the theoretical model. \S \ref{s:pert} describes solutions obtained using perturbation methods for viscous drag. \S \ref{s:num_mod} compares the analytical solutions thus obtained against numerical solutions of the model. The numerical solutions are also used to compare model predictions of viscous and non-viscous drag. Conclusions are drawn in \S \ref{s:conlusion}.

\section{Mathematical model}\label{s:maths}

\subsection{Equation of motion of a floating object}\label{s:EOM_der}
The motion of a floating inertial object is described by Newton's second law:
\begin{equation}
    m\dot{\mathbf{v}}= \mathbf{F}\equiv\mathbf{B}+\mathbf{M}+\mathbf{G}+\mathbf{R}  \text{,} \label{eq:EOMXZ}
\end{equation}
where $m$ is the mass of the object and $\mathbf{v}$ its velocity with the dot denoting a derivative with respect to time. The total force on the object $\mathbf{F}$ can be decomposed into a buoyancy force $\mathbf{B}$, an added-mass force $\mathbf{M}$, a gravity force $\mathbf{G}$ and a resistance force $\mathbf{R}$, which are formulated below. The buoyancy and added-mass forces arise from the integral of pressure around the object. For simplicity, we will assume the object is spherical with diameter $D$. Throughout, it is assumed that the object is small relative to the wavelength, such that $D/\lambda_0\ll 1$, with $D$ the diameter of the object and $\lambda_0$ the wavelength. This has four important consequences. First, the wave field is unaffected by the presence of the object; in other words, there is no diffraction. Second, the free surface can be approximated as an (inclined) straight line on the scale of the object. Third, we can approximate the (relative) velocity field between the liquid and object, which determines the drag on the object, as the velocity at a point. Fourth, the buoyancy force can be computed from the submergence measured relative to the free surface. Nevertheless, the model neglects surface tension. This assumption is reasonable for floating objects provided the following threshold criterion (e.g \cite{falkovich2005floater}) is met: $D/2 > \sqrt{\gamma/(\rho g)}$, where $\gamma$ is surface tension, $\rho$ is density of water and $g$ is gravitational acceleration. For water, the criterion is satisfied for objects of diameter exceeding 5.4 mm, resulting in the findings being invalid for microplastic. However, such small plastics are likely to behave as purely Lagrangian tracers.

\begin{figure}
\centering
\includegraphics[width=\textwidth]{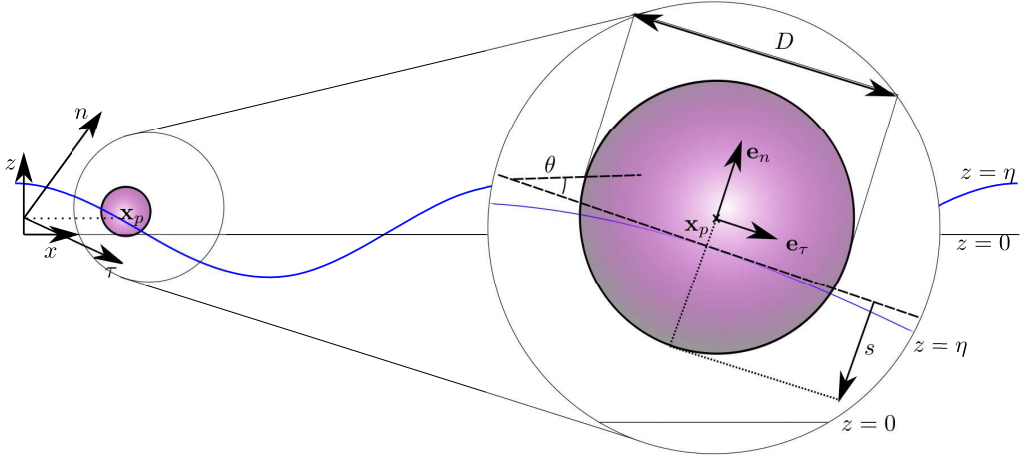}
\caption{Diagram of the two coordinate systems used to describe a floating object of diameter $D$: a stationary laboratory coordinate system ($x$, $z$) and a vertically translating and rotating coordinate system ($\tau$, $n$) with its origin at the vertical position of the free surface $z=\eta_p$ and the $\tau$-axis aligned tangential to the free surface. The vector $\mathbf{x}_p$ locates the centre of the object relative to the origin of the stationary coordinate system, $\tan\theta=\partial \eta/\partial x$ is the slope of the free surface, and $s$ is the (variable) submergence.}
\label{fig:object_FS}
\end{figure}

We first adopt a stationary two-dimensional laboratory coordinate system ($x$, $z$) with the vertical coordinate $z$ measured upwards from the undisturbed free surface. To define the forces on the object, a second, moving coordinate system ($\tau$, $n$) is established that moves vertically with the free surface $z=\eta(x,t)$ and aligns locally with the $\tau$-axis tangential to the free surface at the position of the object $x_p$ and the $n$-axis normal to it, as shown in figure \ref{fig:object_FS}. The coordinate transformation takes the form of a vertical translation followed by a clockwise rotation through the angle $\theta=\arctan\left(\partial \eta/\partial x\right)$, both at the horizontal position of the object $x_p$:
\begin{multline}\label{eq:coord_trans1}
\begin{bmatrix} \tau \\ n \end{bmatrix}=
\begin{bmatrix} 1 &  \partial_x \eta(x,t)|_{x_p}\\ -\partial_x \eta(x,t)|_{x_p} & 1 \end{bmatrix}
\begin{bmatrix} x \\ z-\eta(x_p,t) \end{bmatrix}
\Xi(x_p,t)
%\\
\\
\textrm{with} \quad \Xi(x_p,t)\equiv \left(1+\left(\partial_x \eta(x,t)|_{x_p}\right)^2\right)^{-1/2},
\end{multline}
where a small-angle approximation on $\theta$ has been used and $\Xi$ is required for the determinant of the transformation matrix to be unity and thus conserve area. The quantities $\partial_x\eta(x,t)$, $\eta(x,t)$ and $\Xi(x,t)$ are evaluated at the object position $x_p(t)$ and are thus solely functions of time $t$. The coordinate system ($\tau$, $n$) does not translate in the horizontal direction, enabling direct estimation of the object's horizontal drift $\overline{v}_x=\overline{\dot{x}}_p$, where the overbar denotes an average over the wave cycle. The time-dependent unit normal vectors are
\begin{equation}\label{eq:unitvector_rotation}
\mathbf{e}_{\tau} = \left[1, \partial_x \eta(x,t)|_{x_p}\right] \Xi(x_p,t)
\quad  \text{and} \quad
\mathbf{e}_n=\left[-\partial_x \eta(x,t)|_{x_p}, 1\right] \Xi(x_p,t).
\specialnumber{a,b}
\end{equation}
It should be emphasized that ($\tau$, $n$) is an accelerating coordinate system, both in terms of rotation and vertical translation. Inverting \eqref{eq:unitvector_rotation}:
\begin{equation}\label{eq:e_z}
\mathbf{e}_x= \left(\mathbf{e}_{\tau}(t) -\partial_x \eta(x,t)|_{x_p} \mathbf{e}_n(t) \right)\Xi(x_p,t)
\quad  \text{and} \quad
\mathbf{e}_z= \left(\partial_x \eta(x,t)|_{x_p}\mathbf{e}_{\tau}(t) + \mathbf{e}_n(t) \right)\Xi(x_p,t). \specialnumber{a,b}
\end{equation}
For the time-dependent unit normal vectors $\mathbf{e}_\tau(t)$ and $\mathbf{e}_n(t)$:
\begin{equation}\label{eq:de_dt}
\frac{\text{d} \mathbf{e}_{\tau}(t)}{\text{d} t}=\dot{\theta}_p \mathbf{e}_{n}(t) \text{ and } \frac{\text{d} \mathbf{e}_n(t)}{\text{d} t}=-\dot{\theta}_p \mathbf{e}_{\tau} (t)
\textrm{ with }
\dot{\theta}_p=\textrm{d}_{t}(\partial_{x}\eta(x,t)|_{x_p}) \Xi_p^2,
\specialnumber{a,b}
\end{equation}
in which $\theta_p(t)\equiv\theta(x_p(t),t)$, $\Xi_p(t)=\Xi(x_p(t),t)$, and ${\rm d}_t\equiv {\rm d}/{\rm d}t$.

Denoting the position of the object as $\mathbf{x}_p=x_p\mathbf{e}_x+z_p\mathbf{e}_z=\eta_p\mathbf{e}_z+\tau_p\mathbf{e}_\tau+n_p\mathbf{e}_n$ with $\eta_p(t)\equiv\eta(x_p(t),t)$, its velocity may be written as:
\begin{equation}\label{eq:vel_define}
    \mathbf{v}=v_x\mathbf{e}_x +v_z \mathbf{e}_z
    = \left(\dot{\tau}_p-\dot{\theta}_p n_p+\dot{\eta}_p\partial_x \eta|_{x_p}\Xi_p \right)\mathbf{e}_{\tau}+\left(\dot{n}_p+\dot{\theta}_p \tau_p +\dot{\eta}_p\Xi_p\right) \mathbf{e}_n,
\end{equation}
where we have used \eqref{eq:de_dt} for the time derivatives of the unit vectors $\mathbf{e}_\tau$, and $\mathbf{e}_n$, and $\mathbf{e}_z$ was substituted for from (\ref{eq:e_z}b). The velocity in the translating reference frame $\mathbf{v}^*$ is related to the velocity in the stationary reference frame $\mathbf{v}$ by $ \mathbf{v}^*=\mathbf{v}-\dot{\eta}_p\mathbf{e}_z $, where both vectors can be expressed in any arbitrary set of orthogonal components, such as $\mathbf{e}_x$ and $\mathbf{e}_z$ or $\mathbf{e}_\tau$ and $\mathbf{e}_n$. Accordingly, the acceleration of the object can be written as:
\begin{eqnarray}
\dot{\mathbf{v}}=\dot{v}_x\mathbf{e}_x +\dot{v}_z \mathbf{e}_z&=&
\left(\ddot{\tau}_p-\ddot{\theta}_p n_p-2\dot{\theta}_p \dot{n}_p-(\dot{\theta}_p)^2\tau_p+\ddot{\eta}_p\partial_x \eta|_{x_p}\Xi_p\right)\mathbf{e}_\tau\nonumber\\
&+&\left(\ddot{n}_p+\ddot{\theta}_p \tau_p+2\dot{\theta}_p \dot{\tau}_p-(\dot{\theta}_p)^2 n_p +\Ddot{\eta}_p\Xi_p\right )\mathbf{e}_n.
\label{eq:v_ddot}
\end{eqnarray}
To evaluate \eqref{eq:v_ddot}, the double time derivatives $\ddot{\theta}_p$ and $\ddot{\eta}_p$ must be evaluated explicitly. The double time derivative $\ddot{\theta}_p$ can be obtained by differentiating with respect to time twice using the relationship $\theta_p=\arctan\left(\partial \eta/\partial x |_{x_p}\right)$, noting that $x_p$ is a function of time requiring the chain rule, to obtain:
\begin{eqnarray}\label{eq:theta_ddot}
    \ddot{\theta}_p=&\left( \partial_{ttx}\eta|_{x_p}+2\dot{x}_p \partial_{txx} \eta|_{x_p} + \left(\dot{x}_p\right)^2\partial_{xxx}\eta|_{x_p}+\ddot{x}_p\partial_{xx}\eta|_{x_p} \right) \Xi_p^2 \nonumber \\
    &+\left(\partial_{tx}\eta|_{x_p} + \dot{x}_p \partial_{xx}\eta|_{x_p} \right) 2 \Xi_p \dot{\Xi}_p.
\end{eqnarray} 
Similarly, the double time derivative $\ddot{\eta}_p$ takes into account the dependence of the free surface $\eta_p(x_p(t),t)$ on time $t$ and the time-dependent horizontal position $x_p(t)$, which gives through the chain rule after differentiating twice:
\begin{equation}\label{eq:eta_ddot}
    \ddot{\eta}_p=\partial_{tt}\eta|_{x_p} + 2\dot{x}_p \partial_{tx}\eta|_{x_p}+ \left(\dot{x}_p\right)^2\partial_{xx}\eta|_{x_p}  + \ddot{x}_p \partial_x \eta|_{x_p}.
\end{equation}
Substituting \eqref{eq:theta_ddot} and \eqref{eq:eta_ddot} into \eqref{eq:v_ddot} and \eqref{eq:v_ddot} thence into \eqref{eq:EOMXZ} results in two second-order differential equations in the ($n,\tau$) coordinate system, which are explicitly given by \eqref{eq:tau_P_ddot} and \eqref{eq:n_P_ddot} in \cref{S:simutaneous_eq}. These two equations contain three second-order time derivatives, and so a third (kinematic) equation relating the second-order derivatives is required to solve the system. Such an equation can for example be found by taking the dot product of \eqref{eq:v_ddot} and $\mathbf{e}_x$ (see  \eqref{eq:x_P_ddot} in \cref{S:simutaneous_eq}). 

For convenience, we express the normal coordinate of the centre of the object $n_p$ in terms of the submergence depth $s$ (see figure \ref{fig:object_FS}). To do so, we assume that $D/\lambda_0\ll 1$ so that the free surface is a locally straight line with $n$-coordinate $n_s=-\partial_x \eta|_{x_p}x_p\Xi_p$ (using \eqref{eq:coord_trans1}, setting $x=x_p$ and $z=\eta_p$). The submergence depth is then given by $s=D/2-(n_p-n_s)=D/2-n_p-x_p\partial_x\eta|_{x_p}\Xi_p$, where $D$ is the diameter of the object. From \eqref{eq:vel_define}, the following expression is obtained for the horizontal velocity of the object:
\begin{equation}
\dot{x}_p=\left(
\dot{\tau}_p-\dot{\theta}_p\left(n_p+\tau_p\partial_x\eta|_{x_p}\right)
-\dot{n}_p\partial_x\eta|_{x_p}
\right)\Xi_p.
\end{equation}
It should be noted that $\dot{s}=-\dot{n}_p-{\rm d}_t(x_p\partial_x\eta|_{x_p}\Xi_p)$.

\subsubsection{Buoyancy and added mass}
We decompose total pressure $p$ into an undisturbed component $p_{\rm undisturbed}$ and a disturbed component  $p_{\rm disturbed}$ owing to the presence of the object. Assuming an object that is small relative to the wavelength ($D/\lambda_0\ll 1$), the undisturbed pressure varies as
$p_{\rm undisturbed}=\rho_f g(\eta(x,t)-z)$ on the scale of the object with $\rho_f$ the density of the fluid, so that the dynamic free surface boundary condition $p_{\rm undisturbed}(z=\eta)=0$ is satisfied, the variation with depth is hydrostatic, and any depth-dependent variation owing the waves (cf. $\exp(k_0 z)$ with $k_0$ the wavenumber) is ignored. 

The undisturbed pressure integrated around the wetted surface results in a buoyancy force acting in the normal direction to the free surface,

\begin{equation}\label{eq:buoyancy}
B_n(t)=\frac{g m }{\beta} \frac{V_{\text{s}}}{V}\Xi^{-1}_p =
\frac{g m }{\beta}\left(3\left(\frac{s(t)}{D}\right)^2-2\left(\frac{s(t)}{D}\right)^3\right)\Xi^{-1}_p ,
\end{equation}
where $g$ is the gravitational constant, $V_s$ is the submerged and $V$ the total volume of the sphere, and $\beta\equiv \rho_o/ \rho_f$ is the ratio of object to fluid density. By including $\rho_f g\eta(x,t)$ in the undisturbed pressure, we have included the Froude--Krylov force resulting from the waves. 

The disturbed component of pressure leads to added mass terms, as derived by \citet{maxey1983}:
\begin{equation}
    M_\tau = \frac{C_{m,\tau}(s) m}{\beta} (\dot{u}_\tau(\mathbf{\tilde{x}}_p,t)-\dot{v}_\tau) \text{ and }M_n = \frac{C_{m,n}(s)m}{\beta} (\dot{ u}_n(\mathbf{\tilde{x}}_p,t)-\dot{v}_n),  \specialnumber{a,b}
    \label{eq:addedmass}
\end{equation}
where $\bm{C}_m=(C_{m,\tau},C_{m,n})$ is the added mass coefficient, which is deliberately left as an unspecified function of submergence $s(t)$ at this stage of the derivation. 

The small-diameter assumption leaves the vertical location, where we should evaluate the velocity of the surrounding fluid in \eqref{eq:addedmass}, unspecified. We set this location to be at the free surface, $\tilde{\mathbf{x}}_p=(x_p,\eta_p)$.%, where we choose $\tilde{z}_p=\eta_p-(s/2)\Xi_p^{-1}$ as the base case, corresponding to half the submergence depth $s$ below the free surface in the ($n$,$\tau$)-coordinate system.

\subsubsection{Gravity forces}
The gravity force acts in the vertical direction, and has the following components in the moving coordinate system,
\begin{equation}
G_\tau(t)=-m g\partial_x \eta|_{x_p} \Xi_p(t)
 \textrm{ and }
G_n(t)=-m g\Xi_p(t).
\specialnumber{a,b}
\end{equation}
\begin{figure}
\centering
\includegraphics[width=\textwidth]{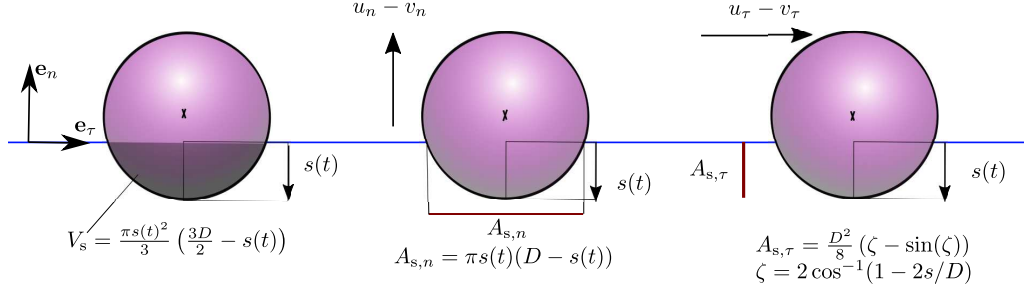}
\caption{Diagrams of (left) the submerged volume $V_{\text{s}}$ as a function of the variable submergence $s(t)$; (centre) the projected area of a submerged sphere moving in the normal direction ($\mathbf{e}_n$); and (right) the projected area of a submerged sphere moving in the tangential direction ($\mathbf{e}_\tau$). All diagrams are shown in the ($\tau$, $n$) coordinate system.}
\label{fig:geomeyry}
\end{figure}
\subsubsection{Resistance forces}
The resistance terms are caused by drag on the object when it has a velocity relative to that of the surrounding liquid. To begin, we assume viscous drag. We assume this drag depends on the submergence of the object and, specifically, we assume the drag is proportional to the submerged projected area of the sphere in the tangential and normal directions (see figure \ref{fig:geomeyry}). Other drag formulations are discussed and examined in \S \ref{s:num_mod}. The resistance force in the tangential direction is,

\begin{equation}\label{eq:stokes_drag}
    R_{\tau}=
    %\frac{1}{2} \rho_f A_{{\rm s},\tau}
     %\frac{24}{\text{Re}_{\tau}} \frac{ \nu \text{Re}_{\tau}}{D } \left( u_{\tau}-v_{\tau}\right)=
      3\pi \rho_f \nu D \hat{A}_{{\rm s},\tau}\left( u^*_{\tau}-v^*_{\tau}\right),
\end{equation}
where $u^*_{\tau}$ and $v^*_{\tau}$ are the velocity components in the $\tau$-direction of the surrounding fluid and the object velocity respectively (in the moving reference frame). The normalised area in the tangential direction $\hat {A}_{{\rm s},\tau}$ is the projected area of the submerged sphere, 

\begin{equation}
A_{{\rm s},\tau}=\frac{D^2}{8}\left(\zeta-\sin(\zeta)\right) 
\quad \textrm{with}\quad 
\zeta\equiv 2\cos^{-1}\left(1-2s/D\right),
\label{eq:Asubtau}
\end{equation}
normalised by the maximum projected area $A= \pi D^2/4$, so that $\hat{A}_{{\rm s}, \tau} = A_{{\rm s},\tau}/A$. Assuming the drag is proportional to the submerged projected area following \citet{beron2016inertia}, which has been validated for steady flows  \citep{miron2020laboratory,olascoaga2020observation}, we evaluate the fluid velocity $u^*_{\tau}$ at the free surface, $\tilde{\mathbf{x}}_p=(x_p,\eta_p)$.

Similar to the $\tau$-direction, we have for the $n$-direction,
\begin{equation}\label{eq:stokes_drag_n}
    R_{n}=3\pi \rho_f \nu D \hat{A}_{{\rm s},n}\left( u^*_{n}-v^*_{n}\right),   
\end{equation}
where we have evaluated the velocity of the surrounding fluid at the same location $\tilde{\mathbf{x}}_p$ as for the tangential resistance force. The submerged projected area of a sphere in the normal direction is given by (see figure \ref{fig:geomeyry}):
\begin{equation}\label{eq:A_s_n}
A_{{\rm s},n}=\pi s(t)\left(D-s(t)\right),
\end{equation}
which again, is normalised by the maximum projected area of a sphere $A= \pi D^2/4$, so that $\hat{A}_{{\rm s},n}= A_{{\rm s},n}/A$. Later, in \S \ref{s:num_mod}, other drag formulations are considered to examine the robustness of the model's predictions.

\subsection{Fluid velocity for surface gravity waves}
We consider unidirectional deep-water surface gravity waves propagating over a horizontal bed in the $(x,z)$-coordinate system, with $z$ measured vertically upwards from still water level, and the free surface located at $z=\eta$. For irrotational flow of inviscid, incompressible  fluid, the governing (Laplace) equation is,
\begin{equation}
\label{eq:laplace}
	 \quad \nabla^2 \phi=0 \quad \text{for } \quad -d \leq z \leq \eta \text{,}
\end{equation}
where $\phi$ is the velocity potential and $d$ depth. Equation \eqref{eq:laplace} is solved subject to the no-flow bottom boundary condition,
\begin{equation}
\label{eq:BBC}
\quad\partial_z \phi=0 \quad \text{for} \quad z=-d \text{,}
\end{equation}
and the kinematic and dynamic linear free surface boundary conditions,
\begin{equation}
\label{eq:KFSBC eta}
u_z-\partial_t \eta- u \partial_x \eta=0 \quad   \text{and} \quad g \eta +\partial_t \phi +\frac{1}{2}(\nabla \phi)^2=0 \quad \text{at} \quad z=\eta \text{,}
\specialnumber{a,b}
\end{equation}
where the velocity components are $u_x=\partial_x \phi$ and $u_z=\partial_z \phi$. 

\section{Perturbation theory for viscous drag}\label{s:pert}
To interpret the physical mechanism behind the drift predicted by the model derived in \S \ref{s:maths}, we use perturbation theory to establish an analytical solution. We do so here for the case of viscous drag, as this allows inclusion of drag at first order in our expansion. We will discuss limitations of viscous drag in \S \ref{sec:limitations_of_viscous_drag} and consider numerical solutions of our model in \S \ref{s:num_mod} in which the assumption of viscous drag is relaxed. We consider only periodic, weakly nonlinear, deep-water surface gravity waves, so that $k_0d\gg 1$ with $k_0$ the wavenumber. We perturb the object position $\mathbf{x}_p$ in a Stokes-type expansion in wave steepness ($\alpha=k_0 a_0$, where $a_0$ the wave amplitude), giving
\begin{equation}\label{eq:x_expand}
    \mathbf{x}_p(t)=\mathbf{x}_p^{(0)} +\alpha \mathbf{x}_p^{(1)}(t)\Big|_{\mathbf{x}_{p}^{(0)}} + \alpha^2 \mathbf{x}^{(2)}_p(t)\Big|_{\mathbf{x}_p^{(0)}}+\mathcal{O}(\alpha^3).
\end{equation}
where the superscript corresponds to the order in $\alpha$, and $\mathbf{x}_p^{(0)}$ is the object label and thus not a function of time. As we are interested in wave-induced drift, which arises at second order, we only pursue those terms necessary to obtain this drift.

Applying a perturbation expansion in the same small parameter $\alpha$ to the governing equation of the fluid \eqref{eq:laplace} and its boundary conditions \eqref{eq:BBC} and \eqref{eq:KFSBC eta} allows the free surface $\eta$ and the velocity potential $\phi$ to be determined, and we do so up to second order.

Although the perturbation theory solutions in this section are for regular waves, the experiments introduced in \cref{s:experiemnts} make use of long (or narrow-bandwidth) wave packets for practical reasons. We assume that inertial effects do not arise on the scale of the packets, as justified in \cref{app:wavepackets}, so that we can correct for the presence of a wave packet simply by accounting for its Eulerian mean flow. \Cref{tab:lin_sol} lists the resulting solutions, whose derivation and laboratory validation is given in more detail by \citet{bremer2019} for deep water and \citet{calvert2019} for intermediate depth. We consider only deep-water waves here ($k_0d\gg 1$). The solutions for the Eulerian return flow and the second-order surface elevation are based on wave packets with envelope $|A_0|$. It is assumed that the wave packets are narrow banded and that the Eulerian return flow is shallow, corresponding to a depth that is small relative to the packet length (\cite{calvert2019} establish the Eulerian return flow without the shallow return flow assumption). In practice, inclusion of the effect of the return flow merely leads to a small correction of less than $2\%$ for our laboratory experiments.

%The Eulerian return flow, which is in the opposite direction to wave propagation, is driven by the set-down of the mean free surface elevation under wavepackets and the divergence of the Stokes drift on the packet scale. Whilst Stokes drift dominates the wave-induced transport near the free surface, the Eulerian return flow is more dominant at depth. 

\begin{table}
    \centering
    \begin{tabular}{lcc}
        Field & Symbol & Solution \\
        \hline
        First-order horizontal velocity & $u_x^{(1)}$ & $A_0 \omega_0 \exp(i \varphi+k_0 z)$ \\
        First-order vertical velocity & $u_z^{(1)}$ &$ -A_0 \omega_0 i\exp(i \varphi+k_0 z)$ \\
        First-order free surface elevation & $\eta^{(1)}$& $A_0 \exp(i \varphi)$ \\
        \hline
        Second-order horizontal Eulerian velocity &  $u_x^{(2)}$ & $-\frac{ \omega_0}{2d}|A_0|^2$ \\
        corresponding time-integrated displacement & $\Delta x_{E}^{(2)}$ & $-\frac{ \omega_0}{2d}\int^{t_2}_{t_1} |A_0|^2 \text{d} t$ \\
        Second-order horizontal Stokes drift velocity & $u_{S}^{(2)}$ & $ k_0 \omega_0 |A_0|^2\exp(2k_0z)$\\ 
        corresponding time-integrated displacement & $\Delta x_{S}^{(2)}$ &  $k_0 \omega_0\int^{t_2}_{t_1} |A_0|^2 \text{d} t$  \\
       % Second-order free surface elevation & $\eta^{(2)}$&$-\frac{1}{4d}|A_0|^2$\\
        \end{tabular}
    \caption{First and second-order solutions for the kinematic properties of deep-water surface gravity waves, with $A_0= a_0\hat{A}_0$ the wave amplitude envelope, $a_0$ its amplitude, $\hat{A}_0$ a non-dimensional envelope, $\omega_0$ the carrier wave frequency, and $k_0$ the carrier wavenumber. Where complex fields are given, the real part is understood, and $\varphi= k_0 x- \omega_0 t$. The first three rows are first-order solutions, valid for regular waves or wave packets. The remaining rows comprise second-order solutions for the wave-averaged Eulerian and Stokes velocities and the set-down. The second-order wave-averaged Eulerian velocity only arises for wave packets, considered in the experiments in \cref{s:experiemnts}.}
    \label{tab:lin_sol}
\end{table}

\subsection{Zeroth-order in wave steepness: $\mathcal{O}(\alpha^0)$}
At zeroth-order in wave steepness, wave forcing evidently does not play a role. Only the normal direction of \eqref{eq:EOMXZ} has any forcing at zeroth order, where the following leading-order static balance is achieved between buoyancy force and gravity,
\begin{equation}
    F^{(0)}_n= \frac{g m}{\beta}\left[ 3 \left(\frac{s^{(0)}}{D} \right)^2- 2\left(\frac{s^{(0)}}{D} \right)^3\right] - gm  =0.
    \label{eq:s0}
\end{equation}
We have used the fact that $\Xi_p=1$ at zeroth order and note that \eqref{eq:s0} is only valid for a floating sphere, i.e. $|D/2-s^{(0)}| \leq D/2$. Equation \eqref{eq:s0} is a cubic equation, which can be readily solved numerically for the depth of submergence of a floating sphere in the absence of waves $s^{(0)}$.

\subsection{First-order in wave steepness: $\mathcal{O}(\alpha^1)$}
We begin by expressing the projected areas of the sphere required to calculate the tangential and normal resistance forces as series expansions around $s^{(0)}$. The submerged projected area of a sphere in the tangential direction \eqref{eq:Asubtau} can be approximated by
\begin{equation}\label{eq:A_sub_expand}
    A_{{\rm s},\tau}(s)=A_{{\rm s},\tau}(s^{(0)})+2D\sqrt{\frac{s^{(0)}}{D}-\left(\frac{s^{(0)}}{D}\right)^2}s^{(1)}+\mathcal{O}(\alpha^2),
\end{equation}
where we have obtained $\partial_s(A_{{\rm s},\tau})$ from \eqref{eq:Asubtau} by implicit differentiation. For the submerged projected area of a sphere in the normal direction, it is sufficient for our purposes to evaluate $A_{{\rm s},n}(s)$ at zeroth order, i.e. $A_{{\rm s},n}(s)=A_{{\rm s},n}(s^{(0)})+\mathcal{O}(\alpha^1)$. 

\subsubsection{The tangential direction}
To first-order of approximation, the velocity and acceleration in the horizontal coordinate $x$ and the tangential coordinate $\tau$ are equal, i.e. $\dot{x}_p^{(1)}=v_x^{(1)}=v_{\tau}^{(1)}$ and $\ddot{x}_p^{(1)}=\dot{v}_x^{(1)}=\dot{v}_{\tau}^{(1)}$. The only forces that play a role are the tangential components of the added mass, gravity and the resistance force. The first-order added-mass terms in the tangential direction are
\begin{equation}
    M_\tau^{(1)}=\frac{C_m m}{\beta}( \dot{u}_x^{(1)} -\ddot{x}_p^{(1)}),
\end{equation}
where we now assume for simplicity that the added-mass coefficient $C_m$ is a constant and independent of direction. Other added-mass formulations are discussed and examined in \S \ref{s:num_mod}.

In a potential flow, a fully submerged sphere has an added mass coefficient of $1/2$. Instead of deriving the complicated dependence of $C_m$ on the object's density, we interpolate linearly between the values for a sphere that is fully submerged ($\beta=1$, $C_m=1/2$) and a sphere that is entirely out of the water ($\beta=0$, $C_m=0$) and set $C_m= \beta/2$. The robustness of this assumption is investigated numerically in \S \ref{s:num_mod}. 

The resistance force \eqref{eq:stokes_drag} can be approximated as:
\begin{equation}
    R_\tau^{(1)}=\Gamma_R m\omega_0 \hat{A}_{{\rm s},\tau}^{(0)}(u_x^{(1)}\vert_{\tilde{\mathbf{x}}_{p}^{(0)}}-\dot{x}_p^{(1)})
    \quad \textrm{with}\quad
    \Gamma_R\equiv \frac{3 \pi \nu D}{\beta V \omega_0},
    \label{eq:R_tau_1st}
\end{equation}
where the non-dimensional coefficient $\Gamma_R$ measures the importance of the resistance force.

From the object's equation of motion \eqref{eq:EOMXZ} we thus obtain:
\begin{equation}
    \left(1+\frac{C_m}{\beta}\right) \ddot{x}_p^{(1)}=\frac{C_m}{\beta} \dot{u}_x^{(1)}|_{\tilde{x}_p^{(0)}} - g  \partial_x \eta^{(1)}\vert_{x_{p}^{(0)}}+\Gamma_R \hat{A}^{(0)}_{{\rm s},\tau}\omega_0 \left(u_x^{(1)}\vert_{\tilde{\mathbf{x}}_{p}^{(0)}}-\dot{x}_p^{(1)}\right).
    \label{eq:ODE_tangential_first_order}
\end{equation}
We seek a solution to the forced second-order ordinary differential equation \eqref{eq:ODE_tangential_first_order} of the form $x_p^{(1)}=\mathcal{R}(i {X}^{(1)}a_0 \exp(i \varphi_{p}^{(0)}))$ with $\varphi_{p}^{(0)}=k_0x^{(0)}_p-\omega_0 t+\varphi_0$ and $\varphi_0=\arg(A_0)$, ignoring initial transients. The complex coefficient $X^{(1)}$ represents the amplitude and phase change of the horizontal motion of the object relative to that of an idealized Lagrangian object under the influence of waves at the same order, $x_{L}^{(1)}=\mathcal{R}(i a_0 \exp(i \varphi_{p}^{(0)}))$. We obtain $X^{(1)}=1$, i.e. there is no horizontal motion amplification compared to that of a Lagrangian particle.

\subsubsection{The normal direction}
Expressing the submergence depth $s$ in terms of the vertical coordinate $z_p$, we have without approximation that $s=D/2-(z_p-\eta_p)\Xi_p$. Therefore, the velocity and acceleration in the vertical coordinate $z$ and the normal coordinate $n$ are related to first order by:
\begin{equation}
\dot{z}_p^{(1)}=v_z^{(1)}=-\dot{s}^{(1)}+\dot{\eta}_p^{(1)}
\quad \textrm{and}\quad
\ddot{z}_p^{(1)}=\dot{v}_z^{(1)}=-\ddot{s}^{(1)}+\ddot{\eta}_p^{(1)}.
    \specialnumber{a,b}
\end{equation}
We first approximate the buoyancy force \eqref{eq:buoyancy} by:
\begin{equation}
B_n^{(1)}= \Gamma_B m\omega_0^2 s^{(1)}
\quad \textrm{with} \quad 
\Gamma_B\equiv \frac{6}{\beta k_0 D}\left(\frac{s^{(0)}}{D}-\left(\frac{s^{(0)}}{D}\right)^2\right),
\label{eq:Bn1}
\end{equation}
the added-mass terms by:
\begin{equation}
    M_n^{(1)}=\frac{C_m m}{\beta} \ddot{s}^{(1)},
    \label{eq:AMn1}
\end{equation}
and the resistance force \eqref{eq:stokes_drag_n} by: 
\begin{equation}
    R_n^{(1)}=\Gamma_R m\omega_0 \hat{A}_{{\rm s},n}^{(0)}\dot{s}^{(1)},
    \label{eq:Rn1}
\end{equation}
where we have used $u_z^{(1)}(z=0)=\dot{\eta}_p^{(1)}$ from the linearised kinematic free surface boundary condition and $v_n^{(1)}=\dot{z}_p^{(1)}$. The new non-dimensional coefficient $\Gamma_B$ measures the strength of dynamic buoyancy, and $\Gamma_R$ measures the strength of the resistance force, as for the tangential resistance force in \eqref{eq:R_tau_1st}. From the object's equation of motion \eqref{eq:EOMXZ} we thus obtain:
\begin{equation} \label{eq:ODE_normal_first_order}
    \left(1+\frac{C_m}{\beta}\right)\left(\ddot{\eta}^{(1)}_p-\ddot{s}^{(1)}\right) = \frac{C_m}{\beta}\dot{u}_z^{(1)}|_{\tilde{x}_p^{(0)}}
    +\Gamma_B \omega_0^2s^{(1)}
    +\Gamma_R \hat{A}^{(0)}_{{\rm s},n}\omega_0 \dot{s}^{(1)},
\end{equation}
where we note gravity only enters at zeroth order. As for the tangential direction,  we seek a solution to the forced second-order ordinary differential equation \eqref{eq:ODE_normal_first_order} of the form $s^{(1)}=\mathcal{R}(\mathcal{S}^{(1)}a_0 \exp(i \varphi_{p}^{(0)}))$ with $\varphi_{p}^{(0)}=k_0x^{(0)}_p-\omega_0 t+\varphi_0$ and $\varphi_0=\arg(A_0)$, ignoring initial transients. We find for the non-dimensional submergence at first order $\mathcal{S}^{(1)}$:
\begin{dmath}\label{eq:L_10_visc}
  \mathcal{S}^{(1)}=\frac{1+\displaystyle \frac{C_m}{\beta}-\Gamma_B-i\Gamma_R \hat{A}^{(0)}_{{\rm s},n}}{\left(1+\displaystyle \frac{C_m}{\beta}-\Gamma_B\right)^2+\left(\Gamma_R \hat{A}^{(0)}_{{\rm s},n}\right)^2}.
\end{dmath}
\begin{figure}
    \centering
    \includegraphics[width=\textwidth]{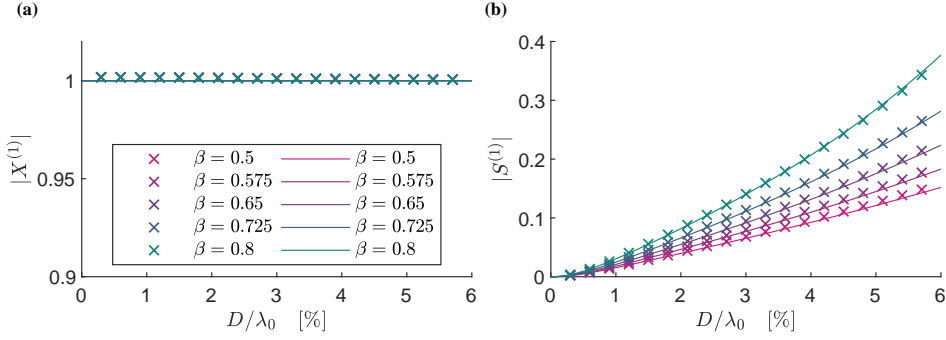}
    \caption{For viscous drag, magnitudes of the first-order horizontal motion amplification $X^{(1)}$ (a) and the variable submergence $\mathcal{S}^{(1)}$ (b) as functions of dimensionless object size $D/\lambda_0$ for different density ratios $\beta=\rho_o/\rho_f$, where the density ratio for each colour is shown in the legend. We have set $C_m= \beta/2$. Numerical and analytical solutions from perturbation theory are denoted by crosses and solid lines, respectively.}
    \label{fig:mag_lin_mot_visc}
\end{figure}
\begin{figure}
    \centering
    \includegraphics[width=\textwidth]{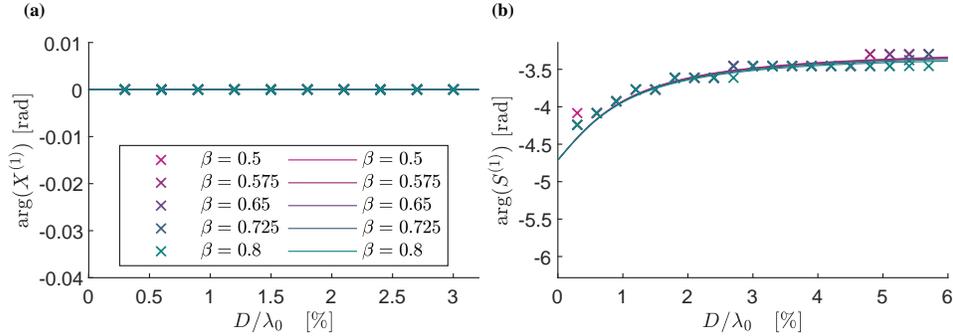}
    \caption{For viscous drag, arguments of the first-order horizontal motion amplification $X^{(1)}$ (a) and the variable submergence $\mathcal{S}^{(1)}$ (b) as functions of dimensionless object size $D/\lambda_0$ for viscous drag and for different density ratios $\beta=\rho_o/\rho_f$, as shown in the legend. We have set $C_m= \beta/2$. Numerical and analytical solutions from perturbation theory are denoted by crosses and solid lines, respectively.}
    \label{fig:arg_lin_mot_visc}
\end{figure}

Figures \ref{fig:mag_lin_mot_visc} and \ref{fig:arg_lin_mot_visc} respectively show the magnitudes and arguments of the first-order solutions for the horizontal motion amplification $X^{(1)}$ and the variable submergence $\mathcal{S}^{(1)}$. In these figures, the purely Lagrangian limit, in which the object is simply transported with the Stokes drift and floats on the moving surface, corresponds to $X^{(1)}=1$, $\mathcal{S}^{(1)}=0$. This limit is obtained as the object size tends to zero. Note that the phase of variable submergence in this limit is non-zero, $\text{arg}(\mathcal{S}^{(1)})\rightarrow \pi/2$. This is because both imaginary and real parts of the variable submergence tend to zero, with the imaginary part approaching zero at a faster rate. As our model is only valid for objects that are small relative to the wave length, we truncate the $x$-axis at $D/\lambda_0=6\%$. Diffraction of the wave field typically only becomes important for $D/\lambda_0>20\%$. %The limits of the analytical solution as the density ratio $\beta$ approaches $0$ or $1$ are discussed in \cref{app:dens_limits}.

As confirmed in figure \ref{fig:mag_lin_mot_visc}a, the magnitude of the horizontal motion $|X^{(1)}|$ is equivalent to that of a purely Lagrangian tracer.  Turning to figure \ref{fig:arg_lin_mot_visc}a, the argument of the horizontal motion $\arg(X^{(1)})$ is evidently also zero. As shown in figure \ref{fig:mag_lin_mot_visc}b, the magnitude of the variable submergence $|\mathcal{S}^{(1)}|$ increases  monotonically with object size and does so at a larger rate for density ratios closer to unity. Variable submergence is driven by the free surface elevation and governed by drag, dynamic buoyancy, and (added) mass, which are respectively the resistance, spring, and inertia terms of a forced spring-mass-damper system (cf. \eqref{eq:ODE_normal_first_order}). The larger the object, the more dominant is the acceleration of the free surface, which acts as an apparent force in the moving reference frame in which the variable submergence is defined, thus increasing the `bobbing' of the object. The lower the density ratio, the stronger the buoyancy force and the stiffer the `spring'. The response in variable submergence for a stiffer `spring' is smaller. The argument of variable submergence $\arg(\mathcal{S}^{(1)})$ decreases monotonically with object size and growing importance of inertia but is dependent on the density ratio, as shown in figure \ref{fig:arg_lin_mot_visc}b.

At first order in steepness the tangential and normal directions are independent, and so it is possible for there to be a significant change in first-order variable submergence whilst the first-order horizontal motion remains unchanged. As can be seen in the next section, a change in first-order variable submergence results in a change in horizontal motion at second order. 

\subsection{Second-order in wave steepness: $\mathcal{O}(\alpha^2)$}\label{S:second_order}
The equation of motion \eqref{eq:EOMXZ} resolved in the horizontal direction and at second order of approximation gives:
\begin{equation}\label{eq:pert_EOM_2}
 \ddot{x}^{(2)}_p= \frac{1}{m}\left(F^{(2)}_{\tau} - \partial_x \eta^{(1)}\Big|_{x_p^{(0)}} F^{(1)}_n\right).
\end{equation}
In order to examine the wave-induced drift of a floating object in periodic waves, we consider the steady wave-averaged transport and set $\overline{\ddot{x}}^{(2)}_p=0$, so that the resultant force must be zero. We will now consider the tangential and normal force contributions to \eqref{eq:pert_EOM_2} in turn.

\subsubsection{Tangential and normal directions}
In the tangential direction, the added-mass terms at second order can be obtained from the combination of an expansion in the horizontal and vertical displacements of the object, a coordinate transformation and evaluation of the advective derivative, respectively:
\begin{dmath}
    M_\tau^{(2)} = \frac{C_m m}{\beta} \left(\dot{u}_x^{(2)}
    +x_p^{(1)}\partial_x \dot{u}_x^{(1)}|_{\tilde{x}_p^{(0)}} +\eta_p^{(1)} \partial_z \dot{u}_x^{(1)}|_{\tilde{x}_p^{(0)}}  
    +\dot{u}_z^{(1)}|_{\tilde{x}_p^{(0)}} \partial_x \eta^{(1)}|_{x_p^{(0)}}
    +\dot{x}_p^{(1)} \partial_x u_x^{(1)}|_{\tilde{x}_p^{(0)}} +\dot{\eta}_p^{(1)} \partial_z u_x^{(1)}|_{\tilde{x}_p^{(0)}}
    -\dot{v}_{\tau}^{(2)}
    \right).
\end{dmath}
In addition to the added-mass terms, the tangential force consists of a correction to the tangential component of gravity due to the object's horizontal displacement,
\begin{equation}
    G_{\tau}^{(2)}=-mg\partial_{xx}\eta^{(1)}\Big|_{x_p^{(0)}}x_{p}^{(1)},
\end{equation}
and a tangential resistance force,
\begin{equation}
    R_{\tau}^{(2)}=3\pi\rho_f\nu D\left(
    \hat{A}_{{\rm s},\tau}^{(1)}\left(u_{\tau,p}^{(1)}-v_{\tau}^{(1)}\right)
    +\hat{A}_{{\rm s},\tau}^{(0)}\left(u_{\tau,p}^{(2)}-v_{\tau}^{(2)}\right)
    \right).
    \label{eq:R_tau2}
\end{equation}
For the first-order velocity components, we have $u_{\tau,p}^{(1)}=u_x^{(1)}|_{\tilde{\mathbf{x}}_p^{(0)}}$ and 
$v_{\tau}^{(1)}=\dot{x}_{p}^{(1)}$. Noting from the coordinate transformation that $u_\tau=u_x+\partial_x\eta|_{x_p}u_z+\mathcal{O}(\alpha^3)$, we obtain for the second-order accurate horizontal fluid velocity component at the object position:
\begin{equation}
    u_{\tau,p}^{(2)}=
    u_{x}^{(2)}|_{\tilde{\mathbf{x}}_p^{(0)}}
    +\partial_x u_{x}^{(1)}|_{\tilde{\mathbf{x}}_p^{(0)}}x_p^{(1)}
    +\partial_z u_{x}^{(1)}|_{\tilde{\mathbf{x}}_p^{(0)}}\tilde{z}_p^{(1)}
    +\partial_x\eta^{(1)}|_{x_p^{(0)}}u_z^{(1)}|_{\tilde{\mathbf{x}}_p^{(0)}}.
    \label{eq:u_tau_p2}
\end{equation}
We set the second-order Eulerian wave-induced velocity $u_{x}^{(2)}$ to zero for the regular waves considered here. The object's horizontal velocity component at second order is:
\begin{equation}
    v_{\tau}^{(2)}=\dot{x}_{p}^{(2)}+\partial_x \eta^{(1)}|_{x_p^{(0)}}\dot{z}_{p}^{(1)},
    \label{eq:v_tau_p2}
\end{equation}
where $\dot{x}_{p}^{(2)}$ is the quantity that is ultimately of interest. Combining \eqref{eq:u_tau_p2} and \eqref{eq:v_tau_p2} and substituting into \eqref{eq:R_tau2} gives:
\begin{dmath}
     R_{\tau}^{(2)}=3\pi\rho_f\nu D\Bigg(
    \hat{A}_{{\rm s},\tau}^{(1)}\left(u_x^{(1)}|_{\tilde{\mathbf{x}}_p^{(0)}}-\dot{x}_{p}^{(1)}\right)
    +\hat{A}_{{\rm s},\tau}^{(0)}\left(
    \partial_x u_{x}^{(1)}|_{\tilde{\mathbf{x}}_p^{(0)}}x_p^{(1)}
    +\partial_z u_{x}^{(1)}|_{\tilde{\mathbf{x}}_p^{(0)}}\eta_p^{(1)}
    -\dot{x}_{p}^{(2)}+\partial_x \eta^{(1)}|_{x_p^{(0)}}\dot{s}^{(1)}\right)
    \Bigg),
    \label{eq:R_tau2subbedin}
\end{dmath}
where we have substituted $u_x^{(2)}=0$, $\dot{z}_{p}^{(1)}=\dot{\eta}_p^{(1)}-\dot{s}^{(1)}$ and $u_z^{(1)}|_{\tilde{\mathbf{x}}_p^{(0)}}=\dot{\eta}_p^{(1)}$ from the linearised kinematic free surface boundary condition. We use the notation $\hat{A}^{(1)}_{{\rm s},\tau}=\hat{A}^{\prime (0)}_{{\rm s},\tau}(s^{(1)}/D)$ with $\hat{A}^{\prime (0)}_{{\rm s},\tau}\equiv \partial_{\hat{s}} \hat{A}_{{\rm s},\tau}(\hat{s})|_{\hat{s}^{(0)}}$ and $\hat{s}\equiv s/D$ according to \eqref{eq:A_sub_expand}.

In the normal direction, the total force at first order consists of a buoyancy force, an added mass and a resistance force already evaluated in \eqref{eq:Bn1}, \eqref{eq:AMn1} and \eqref{eq:Rn1}, respectively. 

\subsubsection{The wave-induced drift}
Substituting the first-order solutions for $x_p^{(1)}$ (i.e. $X^{(1)}=1$) and for $s^{(1)}$ from \eqref{eq:L_10_visc} and for the wave quantities from table \ref{tab:lin_sol} and averaging over the waves, we obtain the following expression from \eqref{eq:pert_EOM_2} for the wave-induced drift of the object $\overline{v}_x=\overline{\dot{x}_p^{(2)}}$:
\begin{dmath}\label{eq:X2}
       \overline{v}_x=\frac{u_{S}}{2}\left[\overbrace{2-\underbrace{\mathcal{R}(\mathcal{S}^{(1)})}_{\substack{\text{Increases}\\\text{drift}}}}^{\text{Adjusted Stokes drift}}
        +\frac{1}{\hat{A}^{(0)}_{{\rm s},\tau}\Gamma_{R}}\left(
    \overbrace{\underbrace{-\Gamma_B \mathcal{I}(\mathcal{S}^{(1)})}_{\text{Increases drift}}}^{\substack{\text{Buoyancy} \\ \text{resolved into} \\\text{the } x\text{-direction}}}
                + \overbrace{\underbrace{
       \frac{C_m \mathcal{I}(\mathcal{S}^{(1)})}{\beta }
         }_{\text{Negligible effect}}}^{\text{Added mass}}
        \right)
         +\overbrace{\underbrace{
         \frac{\hat{A}^{(0)}_{{\rm s},n}}{\hat{A}^{(0)}_{{\rm s},\tau}}\mathcal{R}(\mathcal{S}^{(1)})
        }_{\text{Reduces drift}}}^{\text{Normal drag}}
     \right],
\end{dmath}
where $u_S=k_0\omega_0a_0^2$ is the Stokes drift. We define the drift amplification factor  $X^{(2)}\equiv\overline{v_x}/u_S$, so that $X^{(2)}$ corresponds to the terms inside the square brackets in \eqref{eq:X2} divided by $2$. Equation \eqref{eq:X2} is the main result of this paper, and we will interpret it below. The text above the terms explains their physical origins, and the text below their effect on the wave-induced drift of the object compared to the Stokes drift. 
\begin{figure}
    \centering
    \includegraphics[width=\textwidth]{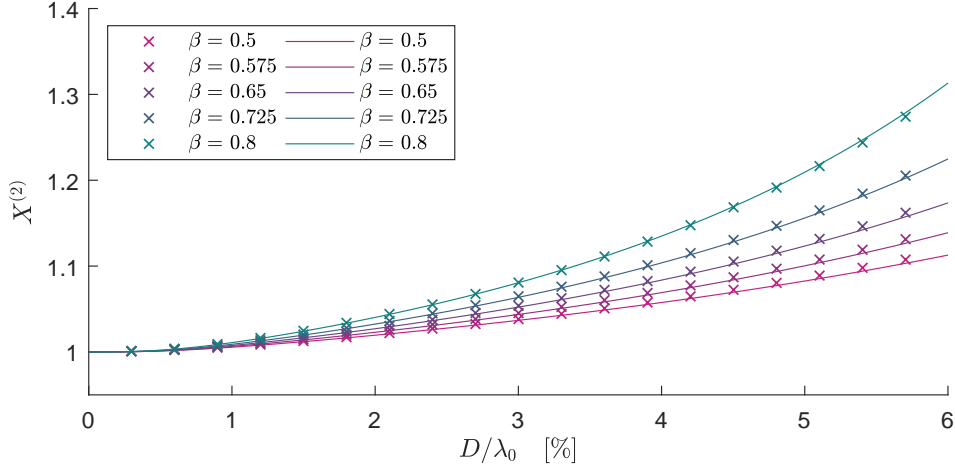}
    \caption{For viscous drag, wave-induced drift amplification $X^{(2)}$ as a function of dimensionless object size $D/\lambda_0$ for different density ratios $\beta=\rho_o/\rho_f$ (see legend). We have set $C_m= \beta/2$. Numerical and analytical solutions from perturbation theory are denoted by crosses and solid lines, respectively.}
    \label{fig:drift_mot_visc}
\end{figure}
\begin{figure}
\centering
\includegraphics[width=\textwidth]{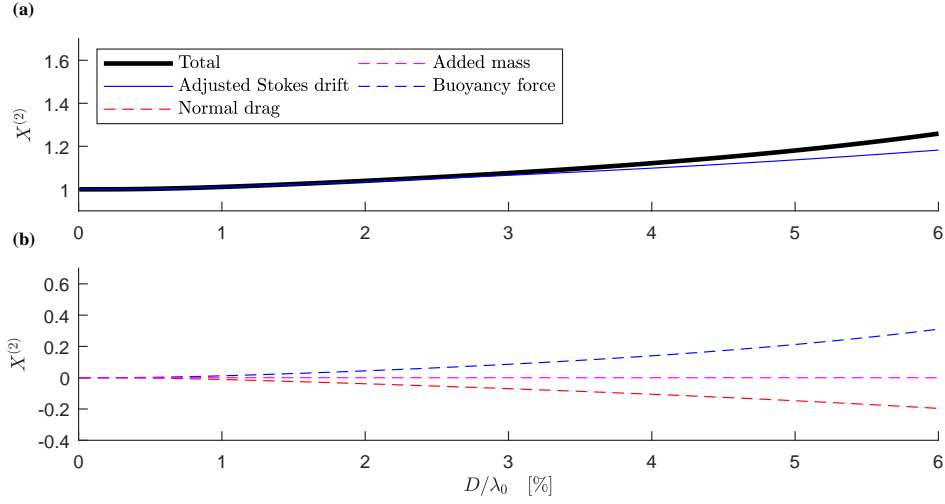}
\caption{For viscous drag, contributions to the wave-induced drift amplification $X^{(2)}$ from the five components in \eqref{eq:X2} as a function of non-dimensional object size $D/\lambda_0$ for density ratio $\beta=0.8$ and $C_m= \beta/2$.}\label{fig:X_2_contri}
\end{figure}

We begin by examining the wave-induced drift amplification factor $X^{(2)}$ as a function of object size and for different density ratios in figure \ref{fig:drift_mot_visc}. It is evident that the drift is enhanced and increasingly so for larger and heavier objects.  Figure \ref{fig:X_2_contri} examines the contributions to $X^{(2)}$ of the four components in \eqref{eq:X2}: the adjusted Stokes drift, buoyancy resolved in the $x$-direction, normal drag, and added mass, which we will discuss in turn. In \eqref{eq:X2} and figure \ref{fig:X_2_contri}, $X^{(2)}=1$ corresponds to objects that do not experience an increase in drift and are simply transported with the Stokes drift (i.e. $\overline{v}_x=u_{S}$). 

\subsubsection{Adjusted Stokes drift}
The adjusted Stokes drift terms in \eqref{eq:X2} reflect change in linear object trajectory. For unmodified horizontal motion ($X^{(1)}=1$) and zero variable submergence ($\mathcal{S}^{(1)}=0$), we obtain $X^{(2)}=1$ from the adjusted Stokes drift terms alone. For larger objects, the increase in the vertical motion due to `bobbing' of the object effectively enhances the Stokes drift,  as shown in figure \ref{fig:X_2_contri}. This mechanism occurs because the linear variable submergence changes the object's orbit and hence its velocity and time spent under trough and crest. Integration of the linear velocity component along the linear orbit results in Stokes drift. Hence, changes to velocity and orbit result in an adjusted Stokes drift. %Note that the contribution of adjusted Stokes drift to the increased wave-induced drift ($1-\mathcal{R}(S^{(1)})$) shown in figure \ref{fig:X_2_contri} becomes $\mathcal{R}(S^{(1)})$ when unity is subtracted from the adjusted Stokes drift curve. 

\subsubsection{Buoyancy resolved in the $x$-direction}
\begin{figure}
\centering
    \begin{tabular}{c|c|c}
        No variable submergence & Variable submergence & Variable submergence \\
         & No normal drag & With normal drag\\
        \hline
         \includegraphics[width=0.3\textwidth]{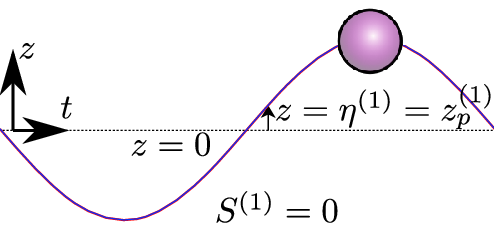}& \includegraphics[width=0.3\textwidth]{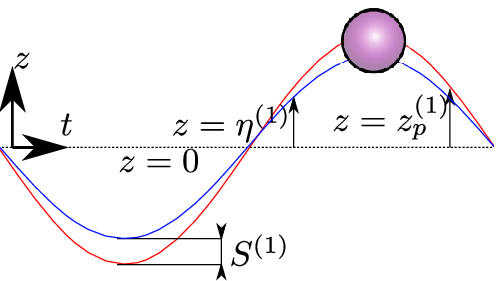}&\includegraphics[width=0.3\textwidth]{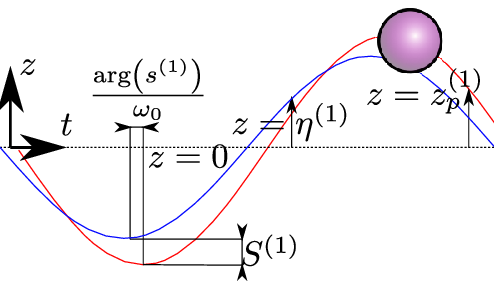}\\
        &&\\
         $\mathcal{S}^{(1)}=0$ &$\mathcal{S}^{(1)} \text{ is real}$ &  $\mathcal{S}^{(1)} \text{ is complex}$ \\
         &&\\
         $\overline{s^{(1)}\partial_x \eta^{(1)}}=0$&   $\overline{s^{(1)}\partial_x \eta^{(1)}}=0$ &
        $\overline{s^{(1)}\partial_x \eta^{(1)}}=A_0^2k_0\frac{\mathcal{I}(\mathcal{S}^{(1)}) }{2}$\\
        &&\\
         &$s^{(1)}$ is out of phase with $\partial_x \eta^{(1)}$.&\makecell{The in-phase component of\\ $s^{(1)}$ with $\partial_x \eta^{(1)}$ has a mean \\ component in the $x$-direction.}\\
         &&\\
          No enhanced drift. &\makecell{No mean component and\\ no enhanced drift.}&\makecell{ This mean component \\ causes an enhanced drift.}\\
    \end{tabular}
    \caption{Schematics of the object trajectory (red) and free surface (blue) for three cases: no variable submergence, variable submergence with no normal drag, and variable submergence with normal drag. The schematics illustrate the physical mechanism for increased drift arising from variable submergence $s^{(1)}$, where variable submergence and drag are in the $n$-direction, and a mean motion in the $x$-direction is created due to the slope of the free surface $\partial_x\eta^{(1)}$. For this illustration, we have chosen a density ratio $\beta=1/2$.}
    \label{tab:mech_drif}
\end{figure}

The mechanism through which buoyancy, when resolved in the $x$-direction and averaged over the wave cycle, can increase the drift of an object is illustrated in figure \ref{tab:mech_drif}. Without variable submergence (left column), the dynamic buoyancy force is simply zero. With variable submergence but without drag in the normal direction (middle column), the first-order buoyancy force resolved in the $x$-direction does not result in a net force on the object, as the first-order buoyancy force and the first-order slope required to resolve this force into the $x$-direction are out of phase. It is only in the presence of a drag component in the normal direction (right column) that a phase lag in the submergence depth arises and a net force results. As shown in figure \ref{fig:X_2_contri}, the buoyancy force thus makes a relatively large contribution to the object's drift. 

\subsubsection{Normal drag}
Although normal drag is required to create the phase difference that leads to the net buoyancy force resolved in the $x$-direction, normal drag also acts to reduce the magnitude of the `bobbing' mechanism and thus reduces the drift motion, as shown in figure \ref{fig:mag_lin_mot_visc}. The horizontal direction component of normal drag opposes the horizontal direction component of buoyancy force, with the balance resulting in a drift that is greater than the adjusted Stokes drift discussed above. Tangential drag, through the inverse dependence of $X^{(2)}$ on the projected area $\hat{A}_{s,\tau}^{(0)}$ and the effective drag coefficient $\Gamma_R$ in \eqref{eq:X2}, acts to reduce the increase in object drift, by effectively `anchoring' the object to the fluid and its Stokes drift.

\subsubsection{Added mass}
At first order, the object accelerates in the normal direction, experiencing an inertia force in addition to the buoyancy force and the normal drag discussed above, and so an added mass term has to be take into account. As shown in figure \ref{fig:mag_lin_mot_visc}, the contribution by added mass is relatively small and acts to reduce drift.

\subsection{Limitation on validity of viscous drag}
\label{sec:limitations_of_viscous_drag}
Although the preceding analysis has demonstrated how enhanced drift of non-infinitesimal objects may arise, the underlying assumption of viscous drag places an upper limit on object size. The maximum Reynolds number that arises from the linear motion in the normal direction is estimated from: 
\begin{equation}
    {\rm Re}_{\rm max} = \frac{a_0 \omega_0\ | \mathcal{S}^{(1)}| D }{\nu} \leq 2,
    \label{eq:Remax}
\end{equation}
where we take $2$ to be the maximum Reynolds number for drag to be considered viscous. Noting that $\mathcal{S}^{(1)}(D/\lambda_0,\beta)$ and taking $\beta=0.8$, we obtain from \eqref{eq:Remax} for the maximum diameter that:
\begin{equation}
\mathcal{S}^{(1)}\left(D_{\rm max}/\lambda_0,\beta=0.8\right)(D/\lambda_0) =\frac{k_0^2\nu }{\alpha\omega_0\pi}
\label{eq:maxRe}
\end{equation}
For a typical laboratory water wave of steepness $\alpha=0.1$ and frequency $f_0=1.25$ Hz, the right-hand side of \eqref{eq:maxRe} becomes equal to $1.6\times10^{-5}$. Fitting a linear curve $S^{(1)}=5.8 D/\lambda_0$ to figure \ref{fig:mag_lin_mot_visc}b, we can solve the quadratic \eqref{eq:maxRe} in $D/\lambda_0$ and obtain a maximum diameter to wavelength ratio of $0.2\%$ corresponding to ${\rm Re}_{\rm max}=2$. Examining figure \ref{fig:drift_mot_visc}, we can conclude that drift enhancement is negligible for such small objects. We will therefore have to use a realistic, non-viscous drag formulation, as discussed in the next section.

\section{Numerical solutions}\label{s:num_mod}
%

%This section describes different formulations of drag and added mass force with reference to \cref{S:sens_anl} which examines the sensitivity of the model to these formulations. Simulations using the non-viscous drag formulation at laboratory scale  and field scale are considered separately below. 

To validate the perturbation theory for viscous drag in \S \ref{s:pert} and to explore the predictions of our model for realistic, non-viscous drag, we set out to obtain numerical solutions of our model. Specifically, we solved the set of differential equations (\ref{eq:tau_P_ddot}-\ref{eq:x_P_ddot}) with the forces described in detail in \S \ref{s:maths} using a numerical ordinary differential equation solver. The fluid velocity and free surface elevation from table \ref{tab:lin_sol} were used as input. We first consider viscous drag in \S \ref{sec:viscous_drag} and then non-viscous drag in \S \ref{sec:nonviscous_drag}, distinguishing conditions (notably Reynolds numbers) that are representative of laboratory (\S \ref{sec:laboratory_scale_results}; see appendix \ref{s:experiemnts} for further details) and field scale (\S \ref{sec:field_scale_results}). \Cref{S:Lim_num_sol} discusses the small-object limit of the numerical solutions. Alternative drag and added-mass formulations are examined in \cref{S:sens_anl}

The numerical solutions commenced from an initial condition in the absence of waves with the object depth set at the static submergence given by numerical solution of \eqref{eq:s0}. Numerical integration in time was carried out using an explicit Runge-Kutta method with variable time step based on \citeauthor{dormand1980family}'s (\citeyear{dormand1980family}) formulation which is fifth order in time and fourth order in accuracy. Avoiding initial transients, wave forcing was ramped up using half of a Gaussian envelope to steady state. A convergence study showed that a Gaussian half width set to 20 wavelengths was sufficient to avoid initial transients, whilst the spatial and temporal convergence were in part resolved by the variable time step method and checked explicitly for the largest objects. Once the object motion reached steady state, its motion components in the $x$ and $z$ directions were effectively linearised using a band-pass filter between $0.8 f_0$ and $1.2 f_0$. The linear phase was determined using the cross-correlation of the linearised object motion and the linearised Eulerian velocity evaluated at the object position in both directions. The object drift velocity was calculated as the gradient of a straight line fitted to the sub-harmonic $x(t)$ motion obtained by low-pass filtering at $0.5f_0$.

\subsection{Viscous drag}
\label{sec:viscous_drag}
The crosses in figures \ref{fig:mag_lin_mot_visc}, \ref{fig:arg_lin_mot_visc} and \ref{fig:drift_mot_visc} display the numerical solutions of the model with a viscous drag formulation for a (small) steepness $\alpha =0.02$. Near perfect agreement is evident with the perturbation theory solutions shown as continuous lines for both the first-order amplitudes (figure \ref{fig:mag_lin_mot_visc}) and phases (figures \ref{fig:arg_lin_mot_visc}) and the second-order drift (figure \ref{fig:drift_mot_visc}). Tiny discrepancies between perturbation theory and numerical simulations in these figures are due to the inherent inclusion of higher-order terms (beyond second-order) in steepness in the numerical simulations. The comparison verifies both the numerical model and the second-order perturbation theory. 
%Figures \ref{fig:all_data} and \ref{fig:X_2_dependance_all} present analytical and  numerical predictions of the dependence of dimensionless linear response and drift factors on object size and density, using the full drag formulation. The small object limit is unity for horizontal motion and zero for variable submergence which corresponds to a purely Lagrangian tracer. The horizontal motion decays with object size, and appears to saturate at a constant value of about 0.95 for non-dimensional object diameter exceeding 5. The reduction in horizontal motion is most evident for smaller objects where inertia is dominant over drag. Both the analytical and predicted estimates of variable submergence linear response are very similar, exhibiting a steady increase with object size, and secondary dependence on density ratio.

%The second-order drift motion initially experiences a very small decrease; this is because horizontal motion decreases faster than variable submergence increases. The horizontal motion then reaches a constant steady state, and drift increases because 'bobbing' continues to grow with object size. The balance between drag terms which reduce drift and the mechanisms for increased drift, described in \cref{s:mech_drift}, favour increased drift with full quadratic drag than with viscous drag. This drives the larger increase in drift motion.

\subsection{Non-viscous drag}
\label{sec:nonviscous_drag}
To overcome the maximum Reynolds-number limit of the viscous drag formulation (of $\text{Re}\equiv |\bm{u}-\bm{v}|D / \nu= 2.5 \times 10^4$), we also consider the following non-viscous drag formulation:
\begin{dmath}
   R_{j}(t)=\frac{1}{2}C_{d}\left(\text{Re}\right) \rho_f A_{{\rm s},j} \left|u^*_{j}(\tilde{\mathbf{x}}_p,t)-v^*_{j}(t)\right|\left(u^*_{j}(\tilde{\mathbf{x}}_p,t)-v^*_{j}(t)\right),\label{eq:exp_drag_t}
\end{dmath}
where the indices $j =n,\tau$ represent the tangential and normal directions; and drag is determined using an experimentally-fitted, non-viscous drag coefficient $C_{d}$. We choose a formulation of the drag coefficient $C_d(\mathbf{\mathbf{\text{Re}}})$ that captures both viscous drag at small Reynolds number, which is linear in velocity difference, and form drag at high Reynolds number. Specifically, we use the fit to experimental data for drag on a sphere obtained by \citet[page 625]{morrison2013}, which is accurate for $\text{Re}<1\times 10^6$:
\begin{equation}\label{eq:C_d}
    C_d(\mathbf{\mathbf{\text{Re}}})=\frac{24}{\mathbf{\mathbf{\text{Re}}}}+ 2.6 \frac{\mathbf{\mathbf{\text{Re}}}/5}{(1+\mathbf{\mathbf{\text{Re}}}/5)^{1.52}}+0.411 \frac{(\mathbf{\mathbf{\text{Re}}}/(2.63\times 10^5))^{-7.94}}{(1+\mathbf{\mathbf{\text{Re}}}/(2.63\times 10^5))^{-8}} +0.25\frac{\mathbf{\mathbf{\text{Re}}}/(1\times 10^6)}{1+\mathbf{\mathbf{\text{Re}}}/(1\times 10^6)},
\end{equation}
where \eqref{eq:C_d} is the same in both directions because the Reynolds number is independent of direction ($\text{Re} \equiv |\bm{u}-\bm{v}|D / \nu$). Taking the small-object and thus the small-Reynolds-number limit of the drag force in \eqref{eq:exp_drag_t} we can recover the viscous drag on a partially submerged sphere \eqref{eq:stokes_drag} and \eqref{eq:stokes_drag_n}. 

% Where laboratory scale is a $1.25$ Hz wave, corresponding to maximum $D/\lambda $ of $6$ \%. Followed by investigation of field-scale simulations to examine the results at higher Reynolds numbers with non-viscous drag. These simulations use a $0.2$ Hz wave with steepness of $\alpha =0.05$, which is typical of a large amplitude wind wave \citep{toffoli2017}, whilst investigating $D/\lambda \leq 4\%$, . 

\subsubsection{Laboratory scale results}
\label{sec:laboratory_scale_results}

\begin{figure}
    \centering
    \includegraphics[width=\textwidth]{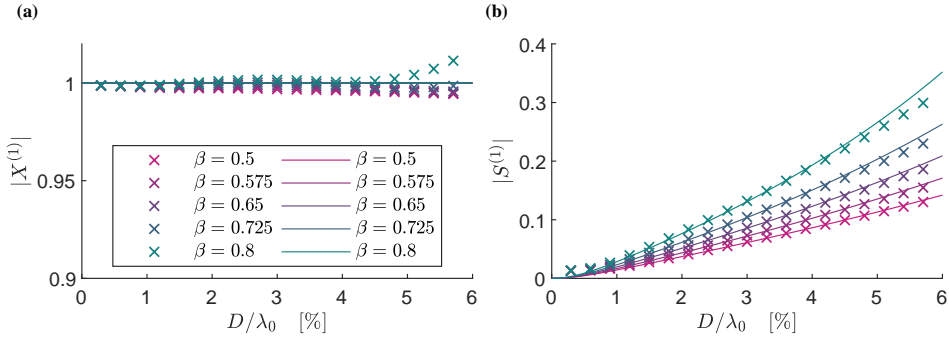}
    \caption{Laboratory scale numerical simulation results using non-viscous drag for magnitudes of the first-order horizontal motion amplification $X^{(1)}$ (a) and the variable submergence $\mathcal{S}^{(1)}$ (b) as functions of dimensionless object size $D/\lambda_0$ for different density ratios $\beta=\rho_o/\rho_f$, where the density ratio corresponding to each colour is listed in the legend. Here, $C_m= \beta/2$. Numerical and analytical solutions from perturbation theory are denoted by crosses and solid lines, respectively.}
    \label{fig:mag_lin_quad_visc_comp}
\end{figure}

At laboratory scale, we set $f_0=1.25$ Hz, corresponding to $\lambda_0=1.0$ m and $\alpha=0.1$. With object diameters up to $D=60$ mm, we obtain $D/\lambda_0=6\%$, where the limit of validity for viscous drag is $D/\lambda_0=0.2\%$ (see \S \ref{sec:limitations_of_viscous_drag}). At laboratory scale, figure \ref{fig:mag_lin_quad_visc_comp} compares the analytically predicted linear motion using viscous drag with the corresponding numerical results using non-viscous drag. The response in the normal direction is unchanged because the forcing is inertial with little effect from drag. As the object size increases, inertia increasingly dominates over drag. A small decrease in horizontal linear motion is evident reaching a few percent for larger objects. The results for small objects are the same because the non-viscous drag recovers viscous drag in the small object limit.

%Although the magnitude of the variable submergence is inertially driven, the phase is dictated by the drag. The phase of the variable submergence decreases when using non-viscous drag, as can be seen in \cref{fig:arg_lin_quad}b, because of the reduction in drag forces. Note that the phase difference is calculated by filtering around the wave forcing frequency of the motion from the numerical simulations and comparing to the linear Eulerian motion at the object position. However, the non-viscous drag changes the spectrum of the response,  creating a small hump to the side of the linear peak in the spectrum of normal motion. This is not an issue when the variable submergence is large enough to dwarf the change, but as object size becomes small, the variable submergence tends to zero. This creates difficulty for filtering and thus comparing the phase, which can be seen for small objects in \cref{fig:arg_lin_quad}b.

%\begin{figure}
    %\centering
    %\includegraphics[width=\textwidth]{arg_lin_quad_v2.eps}
    %\caption{Laboratory scale numerical simulation results using non-viscous drag of arguments of the first-order horizontal motion amplification $X^{(1)}$ (a) and the variable submergence $\mathcal{S}^{(1)}$ (b) as functions of dimensionless object size $D/\lambda_0$ for different density ratios $\beta=\rho_o/\rho_f$, as shown in the legend. We have set $C_m= \beta/2$. Numerical and analytical solutions from perturbation theory are denoted by crosses and solid lines, respectively.}
   % \label{fig:arg_lin_quad}
%\end{figure}

\begin{figure}
    \centering
    \includegraphics[width=\textwidth]{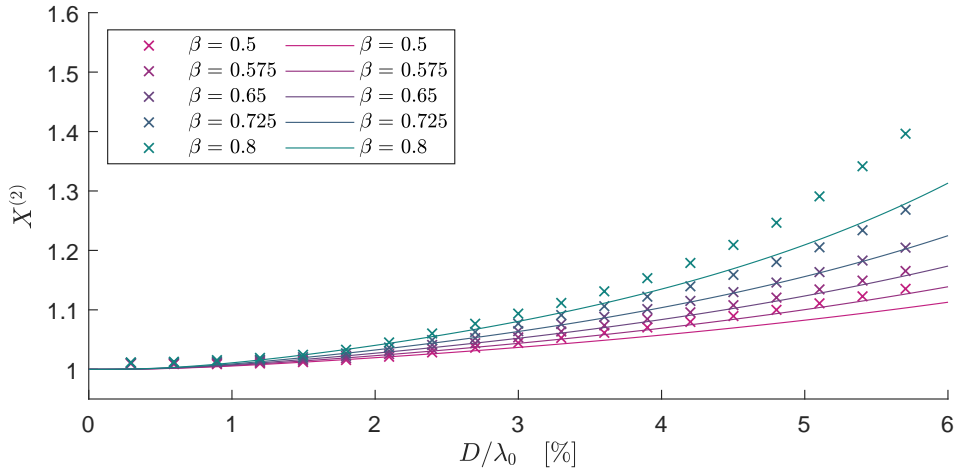}
    \caption{Laboratory scale numerical simulation results using non-viscous drag for wave-induced drift amplification $X^{(2)}$ as a function of dimensionless object size $D/\lambda_0$ and for different density ratios $\beta=\rho_o/\rho_f$ (see legend). Here, $C_m=\beta/2$. Analytical solutions using viscous drag from perturbation theory are denoted by  solid lines.}
    \label{fig:drift_quad_visc_comp}
\end{figure}

The drift amplification increases slightly when using non-viscous drag for larger objects, as seen in figure \ref{fig:drift_quad_visc_comp}. This is because the (tangential) drag force for larger objects is lower for non-viscous drag than for viscous drag, resulting in reduced resistance to increased drift compared to the Stokes drift. The maximum Reynolds number reached in the numerical solutions at laboratory scale was $\text{Re}_{\text{max}}=3.1 \times 10^4$.

\subsubsection{Field scale results}
\label{sec:field_scale_results}
We set a wave frequency of $f_0=0.2$ Hz and a steepness of $\alpha =0.05$ to represent a typical wind wave at field scale. The frequency of $0.2$ Hz corresponds to the peak in the spectrum with $\alpha =0.05$ at the upper end of the steepness range for wind waves in the ocean \citep{toffoli2017}. This steepness corresponds to a dimensional wave amplitude of $a_0=0.3$ m. The difference between viscous and non-viscous drag results will be larger at field scale owing to the higher value of Reynolds numbers, which reached a maximum of $\text{Re}_{\text{max}}= 7.3\times 10^5$ in the numerical simulations.

\begin{figure}
    \centering
    \includegraphics[width=\textwidth]{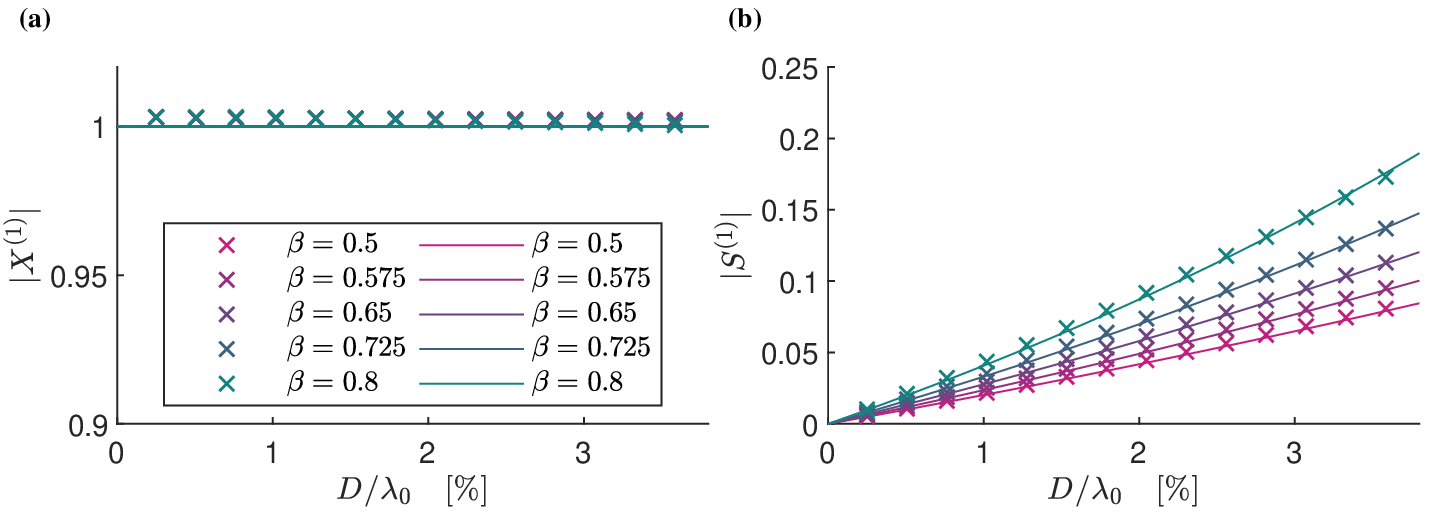}
    \caption{Field scale numerical simulation results using non-viscous drag for magnitudes of the first-order horizontal motion amplification $X^{(1)}$ (a) and variable submergence $\mathcal{S}^{(1)}$ (b) as functions of dimensionless object size $D/\lambda_0$ for different density ratios $\beta=\rho_o/\rho_f$, where the density ratio corresponding to each colour is shown in the legend. Field scale here denotes a $0.2$ Hz wave with a steepness of $\alpha =0.05$. Here, $C_m= \beta/2$. Numerical and analytical solutions from perturbation theory are denoted by crosses and solid lines, respectively.}
    \label{fig:mag_lin_field}
\end{figure}

\Cref{fig:mag_lin_field}a shows the linear horizontal motion, which is mostly unchanged from the perturbation theory result. The magnitude of variable submergence is inertia-driven and thus very similar to the viscous analytical result shown in figure \ref{fig:mag_lin_field}b.

%The phase of the variable submergence significantly decreases at field scale, as can be seen in \cref{fig:arg_field}b, because of the relative reduction in drag when using the non-viscous formulation. %The same difficulty in calculating the phase for small objects due to the non-linear non-viscous drag can be seen in \cref{fig:arg_field}b.

\begin{figure}
    \centering
    \includegraphics[width=\textwidth]{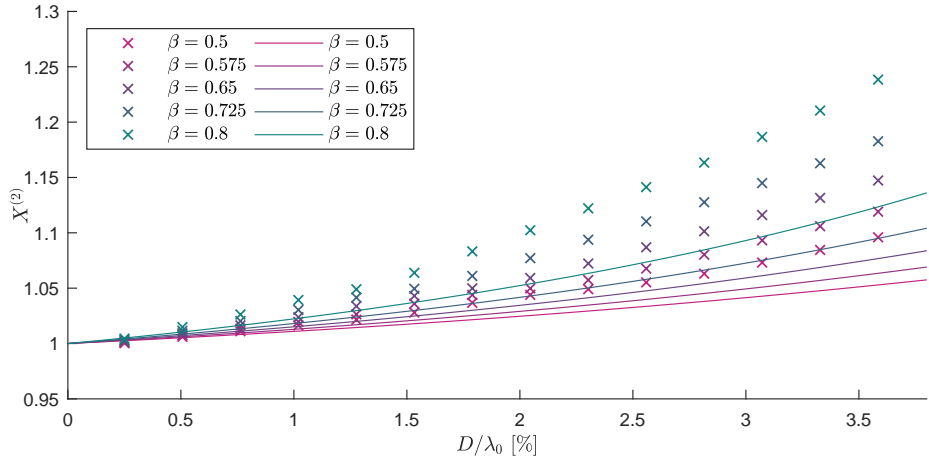}
    \caption{Field scale numerical simulation results using non-viscous drag for the wave-induced drift amplification $X^{(2)}$ as a function of dimensionless object size $D/\lambda_0$ for different density ratios $\beta=\rho_o/\rho_f$ (see legend). Field scale is modelled by a $0.2$ Hz wave with a steepness of $\alpha =0.05$. Here, $C_m=\beta/2$. Analytical solutions using viscous drag from perturbation theory are denoted by  solid lines.}
    \label{fig:drift_quad_field_scale}
\end{figure}

The drift amplification for field scale simulations using non-viscous drag shown in figure \ref{fig:drift_quad_field_scale} is greater than the perturbation theory result based on viscous drag, and even more so than at laboratory scale. This is because the non-viscous drag force is now considerably smaller than its viscous equivalent (taken outside the range of Reynolds numbers for which it is valid). The (tangential) drag force obtained for larger objects is lower for non-viscous drag than for a viscous drag formulation, resulting in reduced resistance to increased drift compared to the Stokes drift.

Using the results from field-scale numerical simulations for non-viscous drag, a $1$ m diameter object of density $\rho_p = 0.9$ $\text{ g}\text{/cm}^{3}$ leads to a $50\%$ increase in drift ($X^{(2)}=1.5$). This is a significant increase compared to the Stokes drift infinitesimal objects would experience. By comparison, a $0.1$ m diameter object in the same wave field does not experience any drift amplification ($X^{(2)}=1$) and behaves as a perfectly Lagrangian tracer. 

\section{Conclusions}\label{s:conlusion}
In this paper, we have developed a model for the transport of spherical, finite-size, floating marine debris by deep-water waves. Using a Stokes-like expansion in wave steepness, we have derived closed-form solutions for the linear response and the wave-induced drift of an object forced by regular waves and experiencing viscous drag. These closed-form solutions match numerical solutions of our model in the case of viscous drag. Our model recovers the Lagrangian limit as object size tends to zero, meaning that small objects are simply transported with the Stokes drift of surface gravity waves.

Through our perturbation solutions, we have identified two mechanisms for increased drift. The first arises from the change in magnitude of the linear orbits, especially its vertical component. The second arises when an out-of-phase variable submergence is resolved in the horizontal direction by the slope of the free surface. The second mechanism requires buoyancy and drag to be acting normal to the free surface, where the drag is required to create the phase difference that gives rise to the drift when averaged over the wave cycle. In any realistic oceanographic scenario, an non-viscous drag is required in order for the drift amplification to be significant. To observe the predicted effect, we have carried out laboratory wave flume experiments for a range of object sizes and densities (see appendix \ref{s:experiemnts}). The experiments show that an increase in wave-induced drift occurs. However, due to large experimental error, the present results have not been used to validate the theoretical model or choice of physics contained within.

The main driver for an increased drift is predicted to be an object's size relative to the wavelength. Thus, in the real ocean, where wavelengths range from $10$-$10^3$ m, increased drift will likely only be observed where shorter wavelengths are present, such as in gulfs or smaller seas. Modelling an object with a diameter of $1$ m and density of $0.9 \text{ g}/\text{cm}^3$ floating on a wave with a $5$ s period and a steepness of $\alpha\equiv k_0 a_0=0.05$, typical of a moderately steep wind wave, results in a $50$\% increase in wave-induced drift compared to the Stokes drift for such a wave. In the same wave field, an object with a diameter of $0.1$ m would not experience an increase in drift at all. High-quality experiments are recommended at larger scale, covering a wider range of object sizes and considering the effect of object shape. Insights from the present work should be useful in the development of more sophisticated models for tracking floating marine litter.

\section*{Acknowledgement}
TSvdB acknowledges a Royal Academy of Engineering Research Fellowship.
\section*{Declaration of interests} 
The authors report no conflict of interest.
\appendix

\section{Equations of motion}\label{S:simutaneous_eq}
Substituting \eqref{eq:theta_ddot} and \eqref{eq:eta_ddot} into \eqref{eq:v_ddot}, and \eqref{eq:v_ddot} into \eqref{eq:EOMXZ} results in two second-order differential equations in the ($n,\tau$) coordinate system:
\begin{dmath}\label{eq:tau_P_ddot}
    \ddot{\tau}_p -\left( -(\partial_x\eta|_{x_p})^2 \Xi_p + \partial_{xx}\eta|_{x_p} \Xi_p^2 n_p\right) \ddot{x}_p=  
      \frac{1}{m(1+\frac{C_{m,\tau}}{\beta})}
    F_\tau
    +\left\{2 \dot{\theta}_p\dot{n}_p +(\dot{\theta}_p)^2 \tau_p \\
    -\partial_x \eta|_{x_p} \Xi_p\left(\partial_{tt} \eta|_{x_p} +2 \dot{x}_p \partial_{tx}\eta|_{x_p} +(\dot{x}_p)^2 \partial_{xx}\eta|_{x_p} \right)
    +n_p\left[ \left(\partial_{tx}\eta|_{x_p} +\dot{x}_p \partial_{xx} \eta|_{x_p}\right) 2 \Xi_p \dot{\Xi}_p\\
    +\left( \partial_{ttx}\eta|_{x_p} + 2\dot{x}_p \partial_{txx} \eta|_{x_p}
    +(\dot{x}_p)^2\partial_{xxx}\eta|_{x_p}  \right) \Xi_p^2 \right]\right\},
\end{dmath}
\begin{dmath}\label{eq:n_P_ddot}
    \ddot{n}_p +\left( \partial_x\eta|_{x_p} \Xi_p + \partial_{xx}\eta|_{x_p} \Xi_p^2 \tau_p\right) \ddot{x}_p=  
    \frac{1}{m(1+\frac{C_{m,n}}{\beta})}F_n
    -\left\{2 \dot{\theta}_p\dot{\tau}_p -(\dot{\theta}_p)^2 n_p \\
    + \Xi_p\left(\partial_{tt} \eta|_{x_p} +2 \dot{x}_p \partial_{tx}\eta|_{x_p} +(\dot{x}_p)^2 \partial_{xx}\eta|_{x_p} \right)
    +\tau_p\left[ \left(\partial_{tx}\eta|_{x_p} +\dot{x}_p \partial_{xx} \eta|_{x_p}\right) 2 \Xi_p \dot{\Xi}_p\\
    -\left( \partial_{ttx}\eta|_{x_p} + 2\dot{x}_p \partial_{txx} \eta|_{x_p}
    +(\dot{x}_p)^2\partial_{xxx}\eta|_{x_p}  \right) \Xi_p^2 \right] \right\}\text{,}
\end{dmath}
where we have kept all the second-order time derivatives on the left-hand side. We now have two equations in terms of three second-order time derivatives, namely $\ddot{\tau}_p$, $\ddot{n}_p$ and $\ddot{x}_p$, and require a third equation to solve the system. We obtain this third (kinematic) equation by taking the dot product of \eqref{eq:v_ddot}, in which we have substituted for $\ddot{\theta}_p$ and $\ddot{\eta}_p$ from \eqref{eq:theta_ddot} and \eqref{eq:eta_ddot}, and $\mathbf{e}_x$, giving:
\begin{dmath}\label{eq:x_P_ddot}
    \ddot{x}_p \left[ 1 +\partial_{xx}\eta|_{x_p} \Xi_p^3 \left( n_p+\partial_x \eta|_{x_p} \tau_p \right) \right] - \ddot{\tau}_p \Xi_p + \ddot{n}_p \partial_x \eta|_{x_p} \Xi_p=\\ %
    \Xi_p\left\{ -n_p\left[ \left(\partial_{tx}\eta|_{x_p} +\dot{x}_p \partial_{xx} \eta|_{x_p}\right) 2 \Xi_p \dot{\Xi}_p
    +\left( \partial_{ttx}\eta|_{x_p} + 2\dot{x}_p \partial_{txx} \eta|_{x_p}\\
    +(\dot{x}_p)^2\partial_{xxx}\eta|_{x_p}  \right) \Xi_p^2 \right] - 2\dot{\theta}_p\dot{n}_p -(\dot{\theta}_p)^2 \tau_p 
    -\partial_x \eta|_{x_p}\left[ \tau_p\left[ \left(\partial_{tx}\eta|_{x_p} +\dot{x}_p \partial_{xx} \eta|_{x_p}\right) 2 \Xi_p \dot{\Xi}_p \\
    +\left( \partial_{ttx}\eta|_{x_p} + 2\dot{x}_p \partial_{txx} \eta|_{x_p}
    +(\dot{x}_p)^2\partial_{xxx}\eta|_{x_p}  \right) \Xi_p^2 \right] +2\dot{\theta}_p \dot{\tau}_p -(\dot{\theta}_p)^2 n_p\right]
    \right\}.
\end{dmath}

\section{Wave flume experiments}\label{s:experiemnts}\label{s:method}

\subsection{Set-up and data acquisition}
\begin{figure*}
\centering
\includegraphics[width=\textwidth]{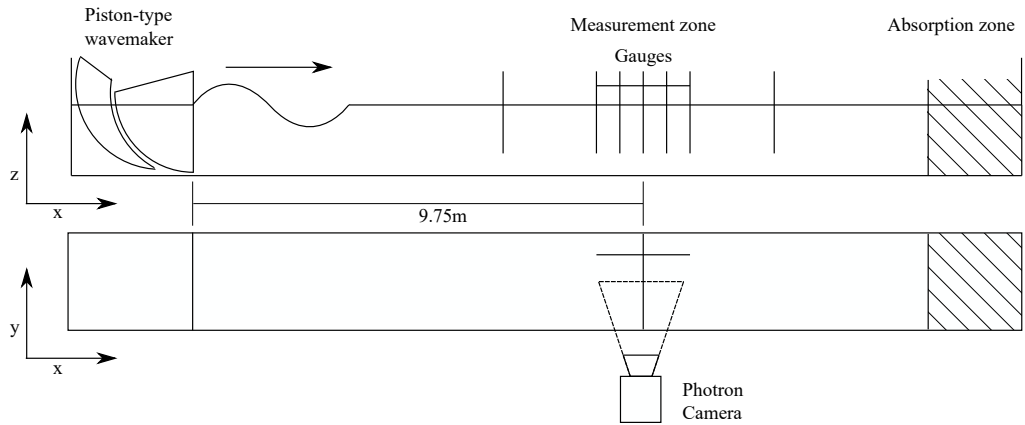}
\caption{Experimental set-up used to track the motion of floating objects under wave motion generated by a double-element piston-type wave maker at the COAST Laboratory,  University of Plymouth, UK.}\label{fig:setup}
\end{figure*}

A series of object tracking experiments were conducted in the Sediment Wave Flume in the Coastal, Ocean and Sediment Transport (COAST) Laboratory at the University of Plymouth, UK. The flume has length $35~\textrm{m}$, width $0.60~\textrm{m}$, and was filled with water to $0.50~\textrm{m}$ depth, as shown in figure \ref{fig:setup}. A double-element piston-type wavemaker supplied by Edinburgh Designs Ltd (EDL) was used to generate a wave packet with a spectral shape that linearly focuses to a Gaussian packet, $A_0=a_0\exp\left(-(x_f-c_{g,0}t)^2/2\sigma^2\right)$, at a measurement zone centred $x_f=9.75~\textrm{m}$  from the rest position of the wavemaker. The wave packet was made as long as possible to make it quasi-monochromatic whilst avoiding reflection ($\epsilon = 1/(k_0 \sigma)=0.04$) with a steepness $\alpha = a_0 k_0=0.1$ and peak frequency $f_0=1.25$ Hz.

Despite our perturbation theory solutions being for periodic waves, we used quasi-monochromatic wave packets in our laboratory experiments because wave-induced transport is much easier to measure experimentally for wave packets (see \cite{vandenbremer_etal2019a} and \cite{calvert2019} and the discussion in \cite{monismith2020}). In appendix \ref{app:wavepackets}, we confirm that the slow modulation associated with the wave packet does not result in any additional non-inertial behaviour of the object. As a result, our model predictions for periodic waves and the wave packets considered in our experiments are equivalent.

We controlled the wavemaker using linear wave theory. Although sub-harmonic error waves at second order generated for wave packets (e.g. \cite{nielsen_baldock2010,orszaghova2014}) can lead to spurious wave-induced displacements \citep{calvert2019}, these displacements are negligibly small for the deep-water waves we consider \citep{vandenbremer_etal2019a}.

Seven resistance-type wave gauges provided $128~\textrm{Hz}$ free surface elevation measurements. Five gauges were located close to the focus location at $15$ cm intervals, as shown in figure \ref{fig:setup} . Two gauges were located significant distances before and after the focus location. After propagating through the measurement zone, the dispersed  wave packets were absorbed by mesh-filled wedges within an absorption zone located at the downstream end of the wave flume. To ensure near-quiescent initial conditions for each experiment, the water surface was allowed to settle for $10~\textrm{minutes}$ between experiments. A Photron SA4 high-speed camera captured the object motions at $125~\textrm{frames/s}$, resolution of $1024$ by $1024~\textrm{pixels}$, and shutter speed of $1/125~\textrm{s}$. Optical distortion was removed using $35~\textrm{mm}$ chequerboard images and MATLAB's inbuilt image processing package.

\subsection{Matrix of experiments}
In the experiments, we selected a peak frequency of $f_0=1.25$ Hz, corresponding to a wavelength of $\lambda_0=1.0$ m and non-dimensional water depth $k_0d=3.1$. We then varied systematically the diameter $D$ and the density $\rho_o$ of the spherical floating object, with values for the 16 experiments listed in table \ref{tab:experimental_matrix}. Object size was limited by camera resolution and the MATLAB tracking algorithm. Density was varied by filling hollow spheres with different ratios of epoxy to glass micro-ball filler. Each experiment was repeated five times. %Effort was made to mix evenly the epoxy and filler, and minimise air voids by using plenty of air release holes whilst filling. Minimum density was limited by the maximum amount of filler that could be mixed into the epoxy whilst it was still workable. 

\begin{table}
    \centering
    \begin{tabular}{c|c|c|c|c}
         Experiment & $D$ [m] &  $\rho_o$ [kgm$^{-3}$] & $D /\lambda_0 \quad [\%]$ & $\beta$ [-] \\
         \hline
        1 &0.051 & 508& 5.1 &0.51\\
        2 & 0.051 &551 &5.1 & 0.55\\
        3& 0.051 & 620&5.1 & 0.62  \\
        4 &0.051 &703&5.1 & 070\\
        5 &0.038 & 597&3.8& 0.60\\
        6 & 0.038 & 637 &3.8&0.63\\
        7 & 0.038 & 678 &3.8& 0.68\\
        8 & 0.038 & 750 &3.8& 0.75\\
        9 & 0.025 & 649 &2.5& 0.65\\
        10 & 0.025 & 678 &2.5& 0.68\\
        11 & 0.025 & 700 &2.5& 0.70\\
        12 &0.025 & 809 &2.5& 0.81\\
        13 & 0.019 & 647 &1.9& 0.65\\
        14 & 0.019 & 679 &1.9& 0.68\\
        15 & 0.019 & 654 &1.9& 0.65 \\
        16 & 0.019 &  807 &1.9& 0.81
    \end{tabular}
    \caption{Matrix of experiments listing dimensional object diameter $D$, object density $\rho_o$, non-dimensional object diameter $D/\lambda_0$, and density ratio $\beta = \rho_o /\rho_f$.}
    \label{tab:experimental_matrix}
\end{table}

\subsection{Data processing}
\begin{figure}
\centering
\includegraphics[width=\textwidth]{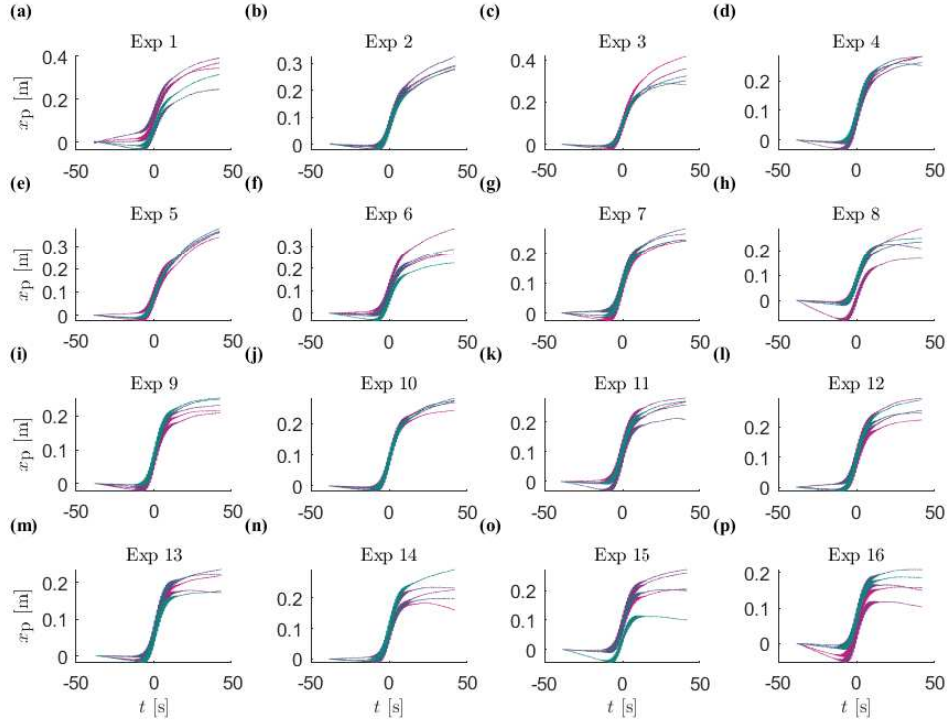}
\caption{Time histories of object horizontal position for each experiment. Each panel shows the five repeated experiments in different colours.}\label{fig:raw_orbits}
\end{figure}
\subsubsection{Free surface elevation}
Wave packets were created from narrow-banded spectra to allow frequency filtering to separate the linear and second-order sub-harmonic components in the wave gauge signal. A band-pass filter between $0.8f_0$ and $1.2f_0$ was used to extract the linear free surface elevation. The measured envelope $A_0$ was calculated using the Hilbert transform of the linear free surface elevation. Use of the measured envelope at the location where the trajectories were measured, to calculate purely Lagrangian displacement, accounted for any dissipation or non-linear dispersion between the wavemaker and the zone of interest.

\subsubsection{Object tracking}
Profile images of the floating white spheres were illuminated from various angles and captured by the Photron camera. The trajectories of the floating objects were tracked by identifying their position in each frame using a circle finding algorithm. The apparent size of the circle in the image was used to calibrate the pixel scale against the known size of the sphere. This also reduced any errors from out-of-plane motion not captured by the single camera. The horizontal components of the raw trajectories, repeated five times, are shown in figure \ref{fig:raw_orbits}. 

Every effort was made to settle the sphere at the start of each experiment in order to give it a zero initial velocity. This was not completely possible due to air flows over the water surface and slight disturbance from human touch. A linear fit in the time domain, assuming a constant pre-existing drift velocity, was used to remove motion before the arrival wave packet from the raw orbits in figure \ref{fig:raw_orbits}. The focus location was determined as coinciding with the position of the maximum of the linearised vertical motion envelope of the object. The difference in object location and exact focus location in the flume had negligible effect because of the very long wave packets used. 

The magnitudes of the linear response were determined by filtering the horizontal and vertical motion components with a band-pass filter of $0.8$-$1.2f_0$, followed by a Hilbert transform to obtain the envelope $A_0$. Note that frequency filtering was only applied to velocities, and numerical integration was used to calculate displacements. The maximum magnitude of the envelope was then normalised by wave amplitude  $a_0$ to obtain $X^{(1)}$ and unity subtracted from the normalised vertical motion to give $\mathcal{S}^{(1)}$ (the normal and vertical directions equivalent up to first-order accuracy). We were not able to extract the linear phase from the experiments because exact spatial and temporal matching of Eulerian wave-gauge data and Lagrangian object positions could not be achieved. A low-pass filter at $0.5 f_0$ was used to extract the sub-harmonic horizontal velocity component. The drift value $X^{(2)}$ was then determined by subtracting the Eulerian return flow from the maximum value of the sub-harmonic horizontal velocity component flow and dividing by the Stokes drift. %Appendix \ref{app:wavepackets} discusses the differences between wavepackets and regular waves and the assumption that the behaviour of the object on the scale of the wavepacket is perfectly Lagrangian, which underlies this procedure.

\subsection{Comparison between theory and experiments} \label{s:results}
\subsubsection{First-order in wave steepness: $\mathcal{O}(\alpha)$}
\begin{figure}
    \centering
     \includegraphics[width=\textwidth]{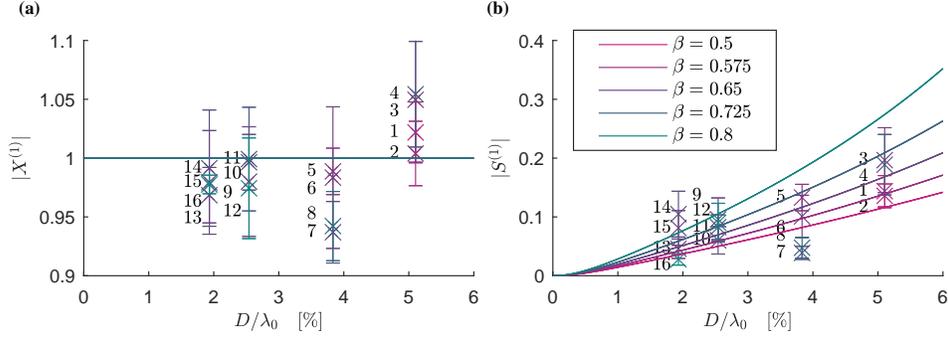}
    \caption{Magnitude of the first-order motion as a function of non-dimensional object size $D/\lambda_0$ for different density ratios (see legend): analytical solution with viscous drag (solid lines) and experiments (circles). The density ratios for the numerical solutions are listed in the legend; density ratios for the experiments are labelled using the same colour scale. The error bars are obtained from repeated experiments and correspond to two standard deviations.}
    \label{fig:all_data}
\end{figure}

\Cref{fig:all_data} presents the first-order magnitudes $|X^{(1)}|$ and $|\mathcal{S}^{(1)}|$ as functions of dimensionless diameter ($D/\lambda_0$) for each experiment, with colour corresponding to density ratio. Comparison is made with numerical solutions of our model for non-viscous drag and analytical solutions using viscous drag. Overall, the horizontal motion in figure \ref{fig:all_data}a is of similar magnitude to what is theoretically predicted ($X^{(1)}$) with some variability, as quantified by the error bars. We note that a decrease of a few percent in the numerical simulation solutions to  $|X^{(1)}|$ is equivalent to a (small) dimensional decrease in the horizontal motion less than $1$ mm. The first-order variable submergence $|\mathcal{S}^{(1)}|$ in figure \ref{fig:all_data}b increases monotonically with dimensionless diameter ($D/\lambda_0$), as predicted by theory. 

The experiments do not show a consistent trend with density for either linear motion component. We note that the densities are not equally spaced or the same for each size sphere owing to practical constraints on filling the spheres with different ratios of epoxy to glass micro-ball filler (see \cref{tab:experimental_matrix} for the experimental matrix). The error bars shown for each experiment, which are twice the standard deviation of the five repeats, are large enough to mask any trend in density. Although we could measure the overall density of the spheres accurately, we emphasize that we were not able to measure its uniformity within the sphere. 

Errors could have arisen from various physical sources that can account for the relatively large standard deviations. The initial motion of the object was hard to eliminate. Air conditioning was switched off, but there were occasional air flows over the flume. The method of taking the value of sub-harmonic velocity at the peak of the wave packet has been shown numerically to match regular waves in \cref{app:wavepackets}. However, inertia at packet scale can be seen in figure \ref{fig:exp_U_sub} as the velocity does not go to zero after the packet passes. Although a 10-minute delay was prescribed between experiments to allow water in the flume to settle, there may have been residual currents still present. The theoretical model also has uncertainty, as can be seen in the sensitivity analysis in \cref{S:sens_anl}, which arises from the choice of drag and added mass formulations, and the exclusion of certain physics from the model, such as surface tension.

\subsubsection{Second-order in wave steepness: $\mathcal{O}(\alpha)$} 
\begin{figure}
\centering
\includegraphics[width=\textwidth]{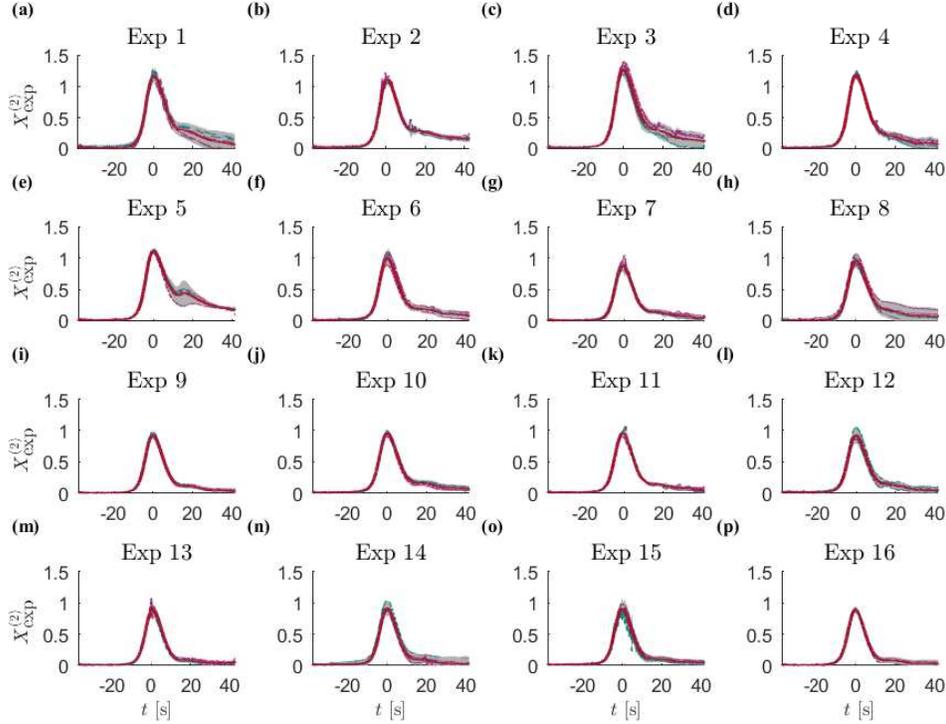}
\caption{Sub-harmonic horizontal object velocity relative to the Eulerian mean flow, normalised by the Stokes drift at the centre of the wave packet: $X^{(2)}_{\text{exp}}=(v_x^{(2)|_{t=0}}-u_x^{(2)}|_{t=0})/(u_{\text{s}}|_{t=0})$ where $u_{\text{s}}|_{t=0}=\omega_0 k_0 a_0^2$. The mean of the five repeated experiments is shown as a continuous red line, and the confidence band corresponding to two standard deviations is shaded in grey, with five lines overlaid for each individual experiment.}
\label{fig:exp_U_sub}
\end{figure}

\begin{figure}
\centering
\includegraphics[width=\textwidth]{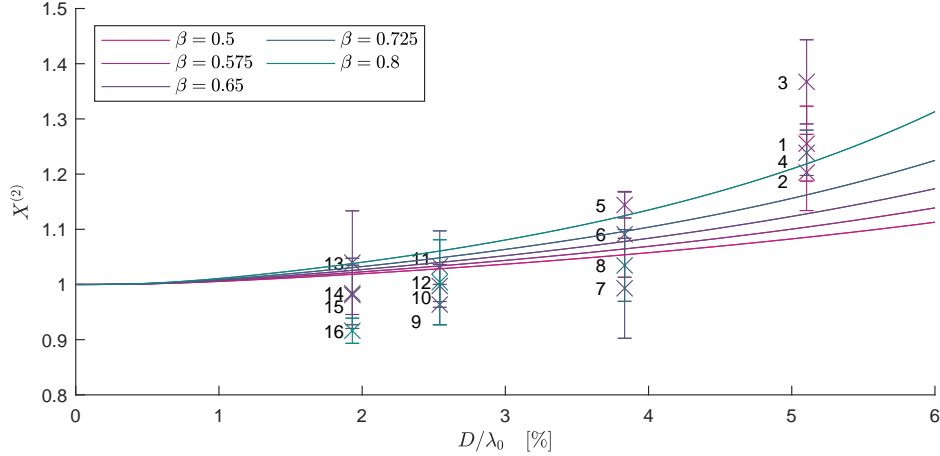}
\caption{Second order drift amplification factor $X^{(2)}$ as a function of non-dimensional object size for different density ratios (see legend): analytical solution with viscous drag (solid lines) and experiments (circles). The density ratios for the numerical solutions are listed in the legend; density ratios for the experiments are labelled using the same colour scale. The error bars are obtained from repeated experiments and correspond to two standard deviations.}\label{fig:X_2_dependance_all}
\end{figure}

\Cref{fig:exp_U_sub} presents time histories of the normalised sub-harmonic horizontal object velocity component for all 16 experiments, having first removed motion ahead of the wave packet and the Eulerian mean flow associated with the wave packet. In all cases, the non-dimensional sub-harmonic horizontal object velocity exceeds or is very close to unity near focus, and has a Gaussian-like profile, reducing close to zero within about 25 s either side of focus.  The distributions are slightly skewed, with a faster rising limb than falling. There is more variability after focus than before. 
Using the peak values from figure \ref{fig:exp_U_sub}, figure \ref{fig:X_2_dependance_all} shows the dimensionless drift factor $X^{(2)}$ for each experiment as a function of dimensionless diameter, with colour indicating density ratio. Drift increases with non-dimensional diameter and, as for the first-order results, the trend with density is unclear from the experiments and masked by substantial variability. We note that the density of floating plastic in the ocean typically has a small range between $800$-$1000$ $\text{kg}/\text{m}^3$ and may thus be a less important variable than object size. The trend with object size is consistent between experiments and theory, both presenting a similar increase with size.

The experiments show that sufficiently large floating objects experience an increase in wave-induced drift. However, the experimental results are not sufficiently accurate to validate the theoretical model. In future work, it is therefore intended to carry out more experiments aimed at validating the model. 

\section{Wavepackets vs. periodic waves}
\label{app:wavepackets}
\begin{figure}
    \centering
    \includegraphics[width=\textwidth]{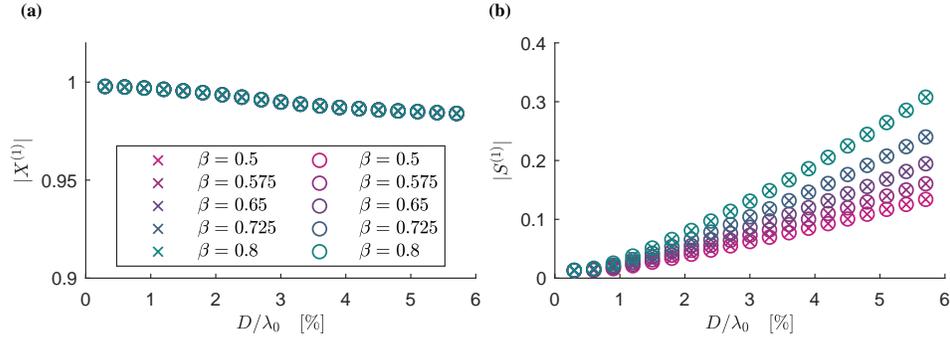}
    \caption{Numerical predictions of the magnitude of the first-order horizontal motion amplification $X^{(1)}$ (a) and the variable submergence $\mathcal{S}^{(1)}$ (b) as functions of dimensionless object size $D/\lambda_0$ for non-viscous drag and for different density ratios $\beta=\rho_o/\rho_f$ (see legend). In the figure, periodic waves are denoted by crosses and  wave packets of the same bandwidth as in the experiments by circles.}
    \label{fig:X_11_WG_reg}
\end{figure}

\begin{figure}
    \centering
    \includegraphics[width=\textwidth]{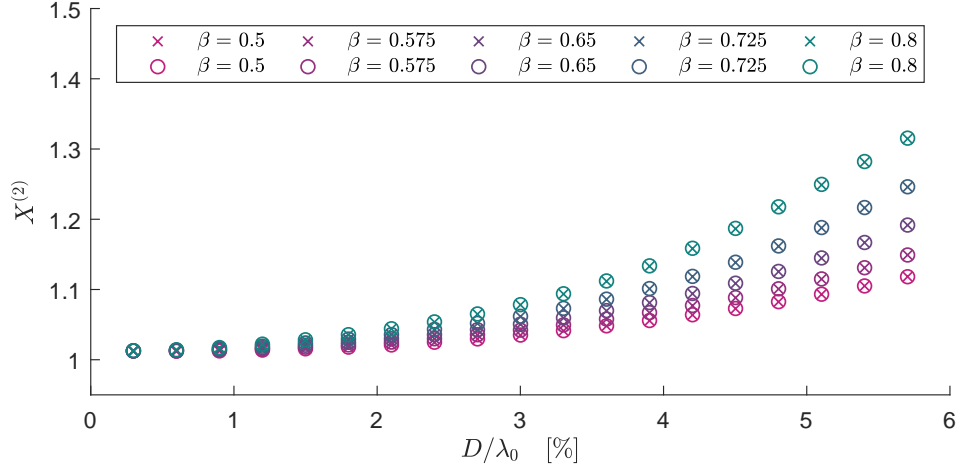}
    \caption{Numerical predictions of the magnitude of the second-order horizontal motion amplification $X^{(2)}$ as functions of dimensionless object size $D/\lambda_0$ for non-viscous drag and for different density ratios $\beta=\rho_o/\rho_f$ (see legend). In the figure, periodic waves are denoted by crosses and  wave packets of the same bandwidth as in the experiments by circles.}
    \label{fig:X_22_WG_reg}
\end{figure}

We use numerical solutions (see \S \ref{s:num_mod}) to the model developed in \S \ref{s:maths} to examine the difference in predictions for objects subject to the quasi-monochromatic wave packets we use in our experiments and periodic waves. The processing of the trajectory data from the numerical simulations using wave packets was the same as for the experiments described in \cref{s:experiemnts}. \Cref{fig:X_11_WG_reg} shows the almost identical first-order response as a function of non-dimensional object diameter at different density ratios for periodic waves (crosses) versus wave packets of the same bandwidth as in experiments (circles). \Cref{fig:X_22_WG_reg} shows the corresponding second-order drift amplification factors. Very slight differences are only predicted for larger object sizes for which the role of inertia is more dominant. For wave packets, a slightly smaller drift motion is predicted, because the time required for inertial objects to reach steady state is longer for larger objects.

\section{Limiting behaviour of the numerical solutions}\label{S:Lim_num_sol}
To confirm the model developed in \S \ref{s:maths} is correct, including its cumbersome coordinate transforms, we examine the perfectly Lagrangian limit (\S \ref{sec:theLagrangianlimit}) and the small-object limit (\S \ref{sec:small_object_limit}) of its numerical solutions obtained using MATLAB's ODE15s solver.
\subsection{The Lagrangian limit}
\label{sec:theLagrangianlimit}
\begin{figure}
    \centering
    \includegraphics[width=\textwidth]{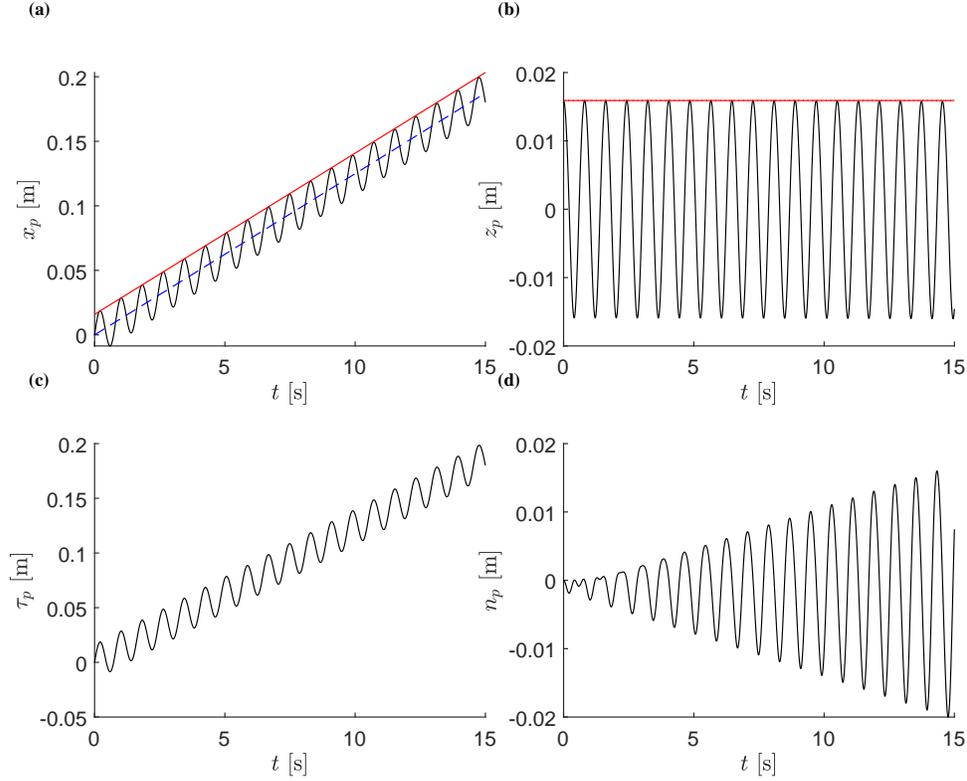}
    \caption{Trajectory of a perfectly Lagrangian tracer obtained using a numerical solution of the present model with forcing provided by \eqref{eq:Lag_tracer}. The top two panels (a, b) display the horizontal and vertical motions $x_p(t)$ and $z_p(t)$, with the blue dashed line showing the theoretical Stokes drift displacement and the red lines the superimposed wave amplitudes. The bottom two panels (c, d) show the tracer particle positions in the ($n,\tau$)-coordinate system.}
    \label{fig:Lag_tracer}
\end{figure}

To obtain the Lagrangian limit, we replace the forces on the object by the accelerations a Lagrangian particle would experience under linear periodic waves:
\begin{equation}
    \ddot{x}_p=a_0 \omega_0^2 \sin(\varphi) \exp\left(k_0 z_p\right), \quad
     \ddot{z}_p=-a_0 \omega_0^2 \cos(\varphi) \exp\left(k_0 z_p\right),
     \specialnumber{a,b}
\label{eq:Lag_tracer}
\end{equation}
where $\varphi = k_0 x_p - \omega_0 t+\varphi_0$. The accelerations are then mapped to the translating coordinate system and expressed in the ($n,\tau$)-directions. The system is then solved numerically in ($n,\tau$)-coordinates and the results mapped back onto ($x,z$)-coordinates, providing confirmation our transformations are correct. As shown in figure \ref{fig:Lag_tracer}, we obtain the correct amplitude of the vertical and horizontal linear motion and the correct Stokes drift.

\subsection{Small-object limit}
\label{sec:small_object_limit}
As object size tends to zero, $D \rightarrow 0$, the solution should recover the behaviour of a perfectly Lagrangian tracer. This has been explicitly checked by numerically solving for an object of non-dimensional diameter $D/\lambda_0= 1 \times 10^{-6}$, which results in $X^{(1)}=1.00$, $\mathcal{S}^{(1)}=0.00$ and $X^{(2)}=1.00$.

\section{Alternative drag and added-mass formulations}\label{S:sens_anl}
This appendix examines several alternative approaches to modelling the drag (\S \ref{sec:drag}) and added-mass (\S \ref{sec:added_mass}) forces on a floating object. Results are obtained from numerical solutions at laboratory scale conditions as in \S \ref{s:num_mod}.

\subsection{Drag}
\label{sec:drag}
Although drag on a fully submerged sphere away from a free surface and in steady flow is well defined across a large range of Reynolds numbers (e.g. \cite{morrison2013}), the drag force on a partially submerged, floating object in the unsteady flow field arising from surface waves is not. To understand the implications for our model's predictions, we consider the following drag formulations: viscous drag with $C_d=24/{\rm Re}$, non-viscous drag with $C_d=C_d(\rm Re)$ based on \cite{morrison2013}, and turbulent drag with $C_d=1/2$. 

\subsubsection{Viscous drag: $C_d=24/{\rm Re}$}
For the viscous drag coefficient $C_d=24/{\rm Re}$, we consider three cases: a case based on submergence-dependent and thus time-varying projected area $\bm{A}_{\rm PA}(t)=(A_{{\rm s},n}(t),A_{{\rm s},\tau}(t))$, as in the paper, a case that ignores the time-dependence and sets $\bm{A}_{\rm PA}=\bm{A}_{\rm PA}^{(0)}\equiv \bm{A}_{\rm PA}(s^{(0)})$, and a case that is based on the time-varying, direction-independent submerged surface area $A_{\rm SA}(t)$. To compute the drag force, we use \eqref{eq:stokes_drag} and \eqref{eq:stokes_drag_n}. For a sphere, the submerged surface area $A_{\text{SA}}(t)= \pi D s(t)$. We normalize this by the surface area of a sphere $A_{\text{FS}}= \pi D^2$, so that $\hat{A}_{\text{SA}}(t)=s(t)/D$ and replace both $\hat{A}_{s,\tau}$ in \eqref{eq:stokes_drag} and $\hat{A}_{s,n}$ in \eqref{eq:stokes_drag_n} by $\hat{A}_{\text{SA}}$. As a result of this normalization, the drag forces on a fully submerged sphere based on projected area and based on submerged area are equal.

The first-order horizontal motion remains unchanged and so is not presented here. Variable submergence and second-order drift solutions are shown in figure \ref{fig:sens_anl_drag}. It is evident that inclusion of time-varying submergence in the projected area and replacing projected by submerged area has a negligible effect on the first-order submergence and only a very minor effect on the drift. 

\subsubsection{Non-viscous drag: $C_d=C_d({\rm Re})$}

For the non-viscous drag coefficient, which is based on a fit to experimental data for a fully submerged sphere \eqref{eq:C_d} (from \cite{morrison2013}), we consider two cases. First, we set the drag to be proportional to the submergence-dependent, time-varying projected area $\bm{A}_{\text{PA}}(t)$, which is the approach used in the paper. Second, we ignore the time dependence and use the projected area of the sphere without waves $\bm{A}_{\rm PA}=\bm{A}_{\rm PA}^{(0)}\equiv \bm{A}_{\rm PA}(s^{(0)})$. 

\begin{figure}
    \centering
     \includegraphics[width=\textwidth]{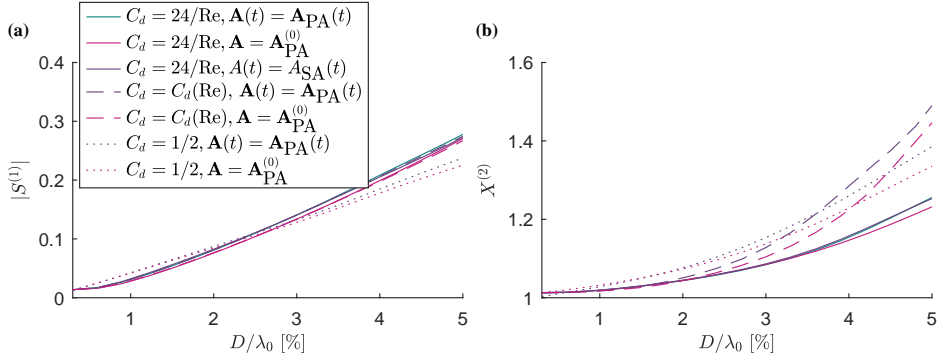}
    \caption{The effect of alternative drag formulations on the numerical predictions of first-order variable submergence $S^{(1)}$ (left) and second-order drift $X^{(2)}$ (right) as a function of non-dimensional object size $D/\lambda_0$ for a density ratio $\beta=0.8$ at laboratory scale conditions. The lines correspond to different drag formulations, labelled in the legend,  using either viscous drag ($C_d= 24/\text{Re}$, solid lines), non-viscous drag ($C_d=C_d(\text{Re})$, dashed lines) or turbulent drag ($C_d=1/2$, dotted lines), which either vary with the time-varying  projected area in the respective directions ($\bm{A}(t)=\bm{A}_{\text{PA}}(t)$), with the constant projected area in the respective directions ($\bm{A}=\bm{A}_{\text{PA}}^{(0)}$), or with the submerged surface area ($ {A}(t)={A}_{\text{SA}}(t)$).}
    \label{fig:sens_anl_drag}
\end{figure}

Again, the first-order horizontal motion is unchanged and not presented here. The magnitude of the variable submergence and the drift are presented in figure \ref{fig:sens_anl_drag}. The variable submergence responses in these two cases are very similar to each other and to the viscous drag cases discussed above.
The solutions for drift are similar to the viscous solution for small objects, diverging as the object size increases. For larger objects, the drift is significantly larger than when modelled with viscous drag. This is caused by the relative reduction in the drag force. There is a slight increase in drift when the projected area is time dependent. 

\subsubsection{Turbulent drag: $C_d=1/2$}
We capture the turbulent-drag limit by setting $C_d=1/2$, which we consider to be the practical large-object limit of \eqref{eq:C_d}. We consider two cases; similar to non-viscous drag, we have used the time-dependent projected areas $\bm{A}_{\rm PA}$ and also consider time-independent projected areas of a sphere in the absence of waves $\bm{A}_{\rm PA}^{(0)}$.

Again, the linear horizontal motion is unchanged and so not presented. The variable submergence is slightly decreased when compared with the viscous and non-viscous cases for larger object sizes, which results in a smaller adjusted Stokes drift. The increase in drift is larger than the viscous cases because of the relative reduction in drag, but smaller than the non-viscous cases. The comparative increase observed when using time-dependent submerged projected area, seen for non-viscous drag, can also be observed with turbulent drag. 

\subsection{Added mass}
\label{sec:added_mass}
\citet{maxey1983} derived the added mass for a fully submerged sphere in a low-Reynolds regime and found the added-mass coefficient to be $C_m=1/2$. \cite{hulme1982} studied a floating hemisphere under wave forcing and derived independent surge and heave added-mass coefficients as functions of non-dimensional object size $k_0 D/2$. The range of non-dimensional object sizes in the present study is  $0 < k_0D/2\le0.16$, which corresponds to added-mass coefficients in the range $0.83 \le C_{m,n}\le 0.86$ in heave and $0.5 \le C_{m,\tau}\le 0.53$ in surge \citep{hulme1982}. 

We consider two categories of added-mass formulations: direction independent and dependent. In the first category, we consider $C_m=0$, $C_m=0.5$ representative of a submerged sphere in a low-Reynolds regime, and $C_m= 0.5 \beta$ for an added mass that increases linearly with depth of submergence in the absence of waves but remains time independent. In the second category, we consider constant $C_m= (0.53,0.83)$ representative of a hemisphere (\cite{hulme1982}), $C_m= 2 \beta (0.53,0.83)$ so that the added mass recovers \citeauthor{hulme1982}'s (\citeyear{hulme1982}) result for a hemisphere and is zero for an entirely unsubmerged sphere. Finally, we extend this to a submergence and time-dependent added mass:  $C_m= 2(0.53,0.83) s(t) /D$.

\begin{figure}
    \centering
    \includegraphics[width=\textwidth]{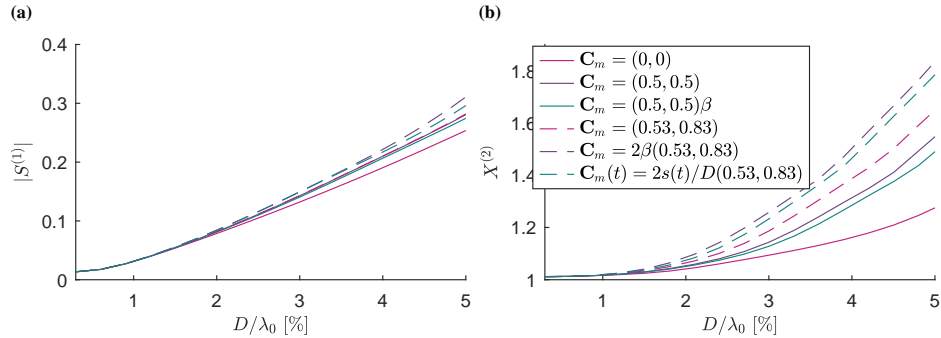}
    \caption{The effect of alternative added-mass formulations on the numerical predictions of first-order variable submergence $S^{(1)}$ (left) and drift $X^{(2)}$ (right) as a function of non-dimensional object size $D/\lambda_0$ for a density ratio $\beta=0.8$ at laboratory scale conditions for non-viscous drag. The lines correspond to different added-mass formulations, described in the legend, with solid lines for directionally independent added-mass formulations, and dashed lines for added-mass formulations decomposed into normal and tangential directions.}
    \label{fig:sens_anl_Cm}
\end{figure}

As for the different drag formulations, the first-order horizontal motion is insensitive to our choice of added-mass formulation. \Cref{fig:sens_anl_Cm} shows the first-order variable submergence and drift responses obtained for the different added-mass formulations considered. The left panel of figure \ref{fig:sens_anl_Cm} shows the relative insensitivity of the variable-submergence response to the different added-mass formulations. The variable submergence exhibits a slight increase when the added mass is directionally dependent and a function of submergence. Drift, shown in the right panel of figure \ref{fig:sens_anl_Cm}, is more sensitive to the choice of added-mass formulation. Direction-independent formulations result in a smaller increase in drift compared to their direction-dependent counterparts. The smallest increase in wave-induced transport (excluding the special case of zero added mass $C_m=0$) is $C_m =0.5\beta$ which is used to generate the analytical and numerical solutions presented in the paper.  

\bibliographystyle{jfm}
\bibliography{Thesis_bib.bib}

\end{document}